\documentclass{article}
\usepackage{lineno,hyperref}
\usepackage{amssymb}
\usepackage{graphicx}
\usepackage{subfig}
\usepackage{siunitx}
\usepackage{url}
\usepackage{booktabs}
\usepackage{multirow}
\usepackage{xcolor}

\begin{document}

\title{Countering Adaptive Network Covert Communication with Dynamic Wardens}

\author{Wojciech Mazurczyk$^1$, Steffen Wendzel$^2$, Mehdi Chourib$^3$, J\"org Keller$^4$\\~\\
\footnotesize $^1$ Institute of Telecommunications, Warsaw University of Technology, Warsaw, Poland\\
\footnotesize $^2$ Department of Computer Science, Worms University of Applied Sciences, Worms, Germany\\
\footnotesize $^3$ Faculty of Mathematics \& Computer Science, FernUniversit\"at in Hagen, Hagen, Germany}
\date{12/01/2018}
\maketitle

\begin{abstract}
\textcolor{red}{This is a preprint. The final version of this paper was published by Elsevier \emph{Future Generation Computer Systems} (FGCS) and can be accessed here: \url{https://doi.org/10.1016/j.future.2018.12.047}.}\\~\\
Network covert channels are hidden communication channels in computer networks. They influence several factors of the cybersecurity economy. For instance, by improving the stealthiness of botnet communications, they aid and preserve the value of darknet botnet sales. Covert channels can also be used to secretly exfiltrate confidential data out of organizations, potentially resulting in loss of market/research advantage. 

Considering the above, efforts are needed to develop effective countermeasures against such threats. Thus in this paper, based on the introduced novel warden taxonomy, 
we present and evaluate a new concept of a \textit{dynamic warden}. Its main novelty lies in the modification of the warden's behavior over time, making it difficult for the adaptive covert communication parties to infer its strategy 
and perform a successful hidden data exchange. Obtained experimental results indicate the effectiveness of the proposed approach.
\end{abstract}


\section{Introduction}\label{Sect:Introduction}
Network information hiding is a discipline that deals with the concealment of network communications or their characteristics and the hidden communication channel created by a data hiding method is referred to as a \textit{network covert channel} \cite{NIHBook}. 


From an economic perspective, covert channels are firstly a contributing factor for malware communications, especially botnet command \& control (C\&C)~\cite{Mazurczyk2015}, \cite{Wendzel2014}. With covert channels, botnets (in comparison to solely applying encryption) can keep their communications secret~\cite{NIHBook,Zander:CCs}. As rental botnets are offered in the darkweb and as covert channels (and encryption) are a major factor in the longevity of such botnet ``products'', their study is a key factor for the success of such botnets. Secondly, another economically important scenario is the fact that covert channels provide a means for the stealthy exfiltration of business data, e.g.,\ internal research results, 
%
%
marketing plans, business contracts or development plans~\cite{NIHBook,Zander:CCs}. Thirdly, covert channel detection, limitation and elimination is a feature of today's anti-malware/-virus software products, which represent an important market. Defeating covert channels is expected to become not only more challenging (as sophisticated techniques emerge) but also more important (as covert channel application increases~\cite{cuing,CACM:CUING}).
From a more general perspective, cyber threats are causing massive economical damage and with the improvement of defense systems, cybercriminals turn to less detectable attack techniques, such as network covert channels~\cite{cuing}. One of the key techniques \textit{ante portas} are \textit{adaptive} covert channels that automatically adjust their covert communication so that they can bypass introduced filter technologies.
Considering these aspects, it is important to develop effective countermeasures to address risks involved with the utilization of network covert channels.

In information hiding the network entity that tries to reveal the existence of (and subsequently eliminate) the covert channel is referred to as a \textit{warden}. A warden's role is to enable auditing, detection, limitation or elimination of such hidden communications \cite{Zander:CCs}, \cite{Fisk:EliminStegInternet}. The warden can be equipped with various countermeasures, which can be classified into passive and active techniques. 

Passive wardens apply non-influential techniques that monitor the traffic (e.g.\ using statistical or behavioral traffic analysis) to detect the involvement of a party in a covert communication or the secret communication itself. On the other hand, active wardens \cite{AndersonPetitcolas:LimitsOfStego} employ measures to monitor, analyze and manipulate the overt traffic (that potentially carries a covert channel), the covert traffic, or both in order to try to disrupt the covert communication (e.g.\ by modifying packet contents, injecting packets into existing connections, inflicting intentional packet losses and/or delays).


An active warden is typically realized by means of traffic normalization. Traffic normalizers aim to remove semantic ambiguities of the protocol header fields or unify the timing behavior of network traffic passing through the normalizer. For example, a normalizer can be equipped with rules to set unused, reserved, and padding bits to zero or to reorder certain portions of packet headers to limit the possibility to successfully hide information there.\footnote{The utilization of reserved/unused header areas and similar approaches is common in covert channel research, cf.\ e.g.~\cite{Zander:CCs,lucena2005covert,wolf1989covert,kundur2003practical}.} Traffic normalization may also affect the passing traffic (both malicious and legitimate) significantly, resulting in the unwanted side-effects.

Primarily, it can have a negative impact on the Quality of Service (QoS) of the passing traffic. The two main sources of this degradation are: \textit{i)} a potentially high number of warden's normalization rules with which the traffic is inspected, which may incur additional delays; \textit{ii)} warden normalization behavior which, for example, can result in adding intentional delays or losses to counter covert channels that exploit the timing aspects of the network traffic. 

Secondly, an important deficit of current wardens (referred to as \textit{regular} wardens in the sequel) is that typically their functioning is static, i.e.,\ the set of active normalization rules typically remains constant in time or changes very rarely and slightly (e.g.\ in reaction to the discovery of some previously unknown data hiding techniques, an administrator may manually add new rules). In particular, this means that if the covert sender and the covert receiver are able to infer the warden's normalization capabilities then they can try to adapt to them \cite{Yarochkin} by adjusting their behavior. This is achieved by first discovering (during a so-called network environment learning (NEL) phase, cf.\ Sect.~\ref{Sect:DynWarden}) and then by exploiting the warden's vulnerabilities, e.g.,\ a lack of certain normalization rules for a particular data hiding method. In result, both covert communication parties can select a network covert channel which is not influenced by the warden in order to perform their hidden communication in an uninterrupted manner.


In this paper, we address the above mentioned deficits, i.e.,\ the warden's static behavior and its impact on the inspected traffic. In particular, we devise a novel warden operation strategy that we call \textit{dynamic}. A \textit{dynamic warden} performs periodic, randomized shuffling of the active subset of normalization rules. In result, the warden is evolving its normalization behavior over time and thus it is increasing the uncertainty for the adaptive covert communication parties by limiting their ability to successfully infer a warden's normalization capabilities.

Considering the above, the main contributions of this paper are to: \textit{(i)} perform a first comprehensive analysis of various aspects and features of a warden which influence its capabilities and effectiveness in countering network covert channels -- to this aim a novel taxonomy for wardens is introduced; \textit{(ii)} propose a novel active warden strategy, i.e.,\ the dynamic warden, which is based on the constantly evolving normalization behavior in order to increase the uncertainty of the covert communication parties and to limit their chance for successful covert communication; \textit{(iii)} create the first proof-of-concept implementation of a dynamic warden; \textit{(iv)} perform an extensive experimental evaluation which proves that this concept is effective and, in particular, allows to achieve a trade-off between the time needed to completely transmit covert data and CPU load, with a longer completion time of the hidden transmission at CPU load similar to a regular warden.

The remainder of this paper is structured as follows. Sect.~\ref{Sect:RelWrk} summarizes the state-of-the-art research on wardens and their capabilities.
In Sect.~\ref{Sect:Taxonomy}, we introduce a novel and comprehensive warden taxonomy which takes various aspects of wardens into account. 
In Sect.~\ref{Sect:DynWarden}, the concept of a dynamic 
warden is introduced which is later evaluated in Sect.~\ref{Sect:Eval}. Sect.~\ref{Sect:Concl} concludes and provides an outlook on future work.

\section{Related Work}\label{Sect:RelWrk}
Investigating the threat caused by covert channels is currently essential in order to assess the cybersecurity of  communication networks in a complete manner. It must be also noted that IP networks with all their services and protocols currently offer a wide range of features that can be utilized for covert channels creation. During the last two decades, many techniques that enable information hiding for various networking environments have been proposed \cite{Zander:CCs}, \cite{Mileva2014}, \cite{Mazurczyk2015a} and continuously they are spanning to the new ICT fields \cite{Wendzel2014}, \cite{Zielinska2014}. It must be also noted that in \cite{Wendzel:CSUR:Patterns} Wendzel et al.\ analyzed more than 100 data hiding techniques that have been grouped into several categories (patterns) based on the way that they function. This illustrates how diverse, complex and sophisticated network covert channels are and therefore this underlines the need for effective countermeasures.

The concept of a warden was first introduced in Simmons' \textit{prisoners' problem} \cite{Simmons:PrisonersProb} where two prisoners, Alice and Bob, are monitored by the jail ward (warden) and try to covertly communicate in order to devise an escape plan. If their secret communication is detected then the prisoners will be confined and thus no longer able to escape. In digital scenarios, the secret communication between Alice and Bob can, e.g., be data hiding in images, audio files or network traffic. In the sequel, we consider network traffic and the warden which is the main entity responsible for inspecting all passing messages in search for the embedded information. Below we review the most notable previous works focused on the warden concept.

Anderson et al.~\cite{AndersonPetitcolas:LimitsOfStego} distinguish between a passive warden, that is capable of monitoring and reporting unauthorized traffic, and an active warden that can modify and drop traffic comprising covert data without unnecessarily interfering a legitimate communication. A subtype of an active warden is the malicious warden \cite{Craver1998} which may act on the legitimate communication with impunity, e.g.,\ by trying to impersonate covert communication parties. A passive warden could use, for example, a network intrusion detection system that monitors and detects irregularities within the passing network traffic \cite{Fisk:EliminStegInternet} or a decision tree- or support vector machine (SVM)-based classifier \cite{zander2011stealthier,Sohn2003,iglesias2017decision} to detect network covert channel utilization.

Most active wardens rely on some kind of traffic normalizer, which is generally a network gateway capable of forwarding, dropping or modifying traffic by, e.g.,\ clearing or resetting a set of bits in the network protocol header fields \cite{Wendzel:Sich12} or by modifying packet delays \cite{WendzelKeller:PAAW}. Several specialized versions of active wardens were introduced, including a number of traffic normalizers such as OpenBSD \textit{pf}'s scrubbing feature \cite{OpenBSD:pf} or the \textit{Snort} normalizer \cite{Snort:norm}.

Another strategy of an active warden is to introduce \textit{random} noise within the detected covert channel \cite{Blasco:FrameworkStego}. As introducing noise can negatively impact legitimate (overt) communications, Fisk et al.\ introduced the concept of a \textit{Minimum Requisite Fidelity} (MRF), referring to an acceptable level of warden-introduced noise for communication peers \cite{Fisk:EliminStegInternet}.


Another type of warden counters \textit{protocol switching covert channels}, i.e.,\ channels that signal hidden information by utilizing different network protocols. Both, detection based on machine learning \cite{PSCC:Detect} and a countermeasure that jumbles packet orders of such protocol switching covert channels \cite{WendzelKeller:PAAW} have been proposed.

An advanced variant of the traffic normalizer is the so-called \textit{Network-Aware Active Warden} (NAAW) which has knowledge of the network topology in which it operates and which implements stateful traffic inspection and normalization. NAAWs were first introduced by Lewandowski et al.\ to normalize potential covert channels in IPv6-based traffic \cite{Lewandowski:NAAW}. NAAWs are based on the concept of active mapping \cite{Shankar:ActiveMapping}, which is a network protection technique that maps the network and its policies to perform improved decisions for traffic normalization.

A detailed description of current techniques used in wardens can be found in~\cite[Chap.~8]{NIHBook}.

%
%

To the best of the authors' knowledge, existing works and systems have the following limitations when compared to the concept introduced in this paper: \textit{(i)} typically, they focus only on a single capability of the warden with respect to the covert communication parties, i.e.,\ most related work focuses on the distinction between passive vs.\ active warden. In this paper, we perform a systematic review of warden types and features; \textit{(ii)} no work investigates relationships between various warden capabilities; \textit{(iii)} most proposed wardens operate in a static manner. If the covert sender and the covert receiver are able to infer the warden's traffic modification strategy then they could adjust their used covert technique and perform their hidden communication uninterrupted.



\section{Warden Types and Features -- a Novel Taxonomy}\label{Sect:Taxonomy}
When considering covert communication, we typically analyze a model in which we distinguish two opposite sides: two (or more) covert communication parties and one (or more) wardens trying to detect, limit or eliminate the hidden data exchange. 

It is also worth noting that, depending on the application or the aim of the covert techniques' utilization, a paradox exists. For some applications, the warden is considered as an adversary (e.g.\ a censor in an oppressive regime) --- in that scenario covert communication parties are trying to protect their privacy and freedom of speech or free access to information. On the other hand, if information hiding techniques are used to exfiltrate user's confidential data, the warden is trying to prevent the actions of an adversary (e.g.\ a malware). Considering the above paradox, we are focusing on the scenario where a warden is a desired security measure whose aim is to fight the malicious utilization of information hiding techniques.

\begin{figure}[!t]
\centering
\includegraphics[width=0.9\textwidth]{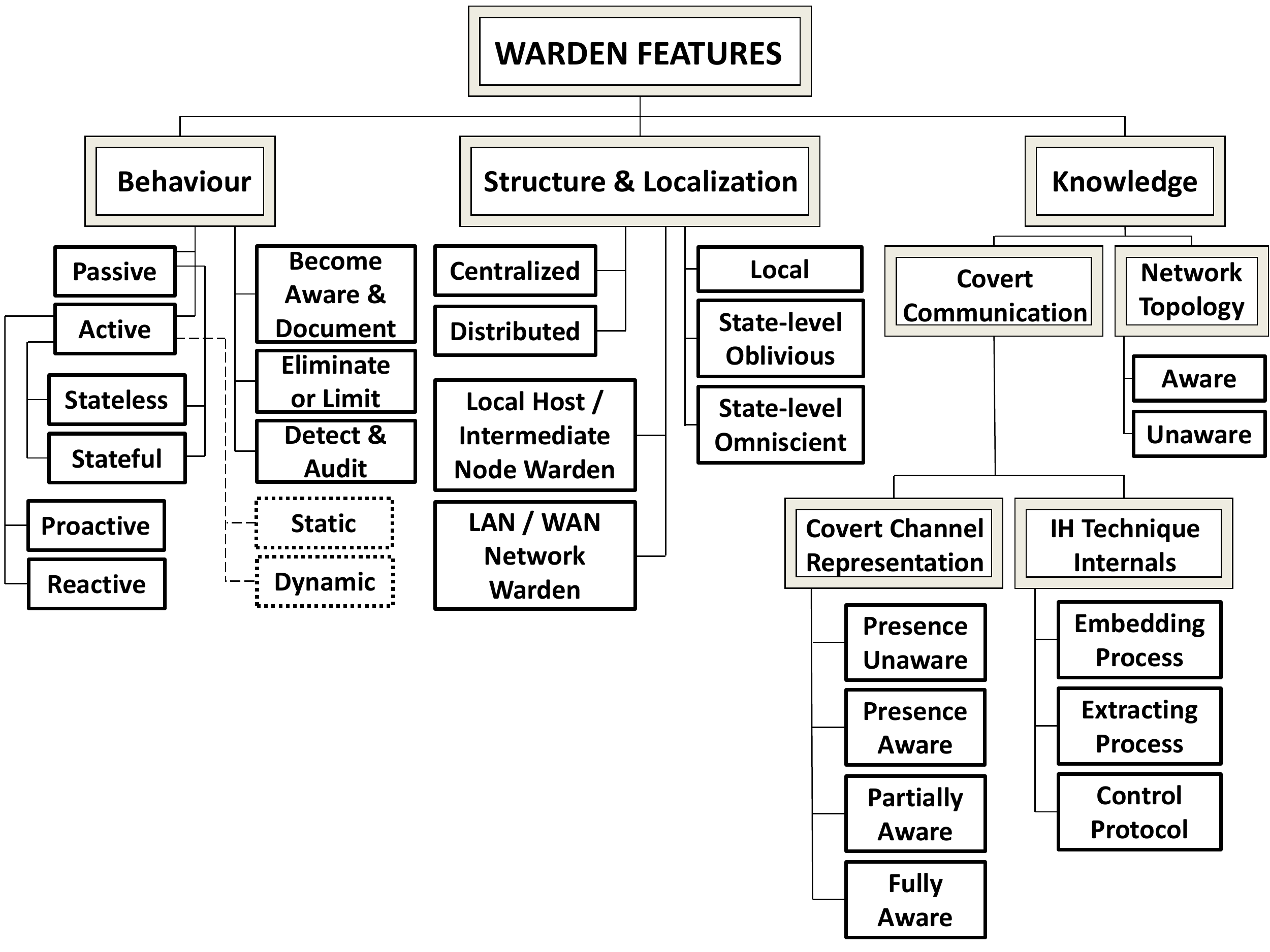}
\caption{The proposed taxonomy of warden features.}
\label{Fig:Taxonomy}
\end{figure}

In the reminder of this section, we present a systematic review on various warden realizations and introduce a new taxonomy depending on the following warden features (Fig.~\ref{Fig:Taxonomy}): behavior, structure, localization, and knowledge about the covert communication process (i.e.\ specifics of the hiding algorithm, how the secret data are presented in the communication channel or on a local host and the hidden data scenario used). To the authors' best knowledge this is the first attempt to provide such a 
comprehensive approach.

\subsection{Warden Structure and Localization}
Depending on the extent of the network infrastructure being controlled by a warden and eventually the administrative body and resources involved, wardens can be divided into: \textit{local, state-level oblivious} and \textit{state-level omniscient} wardens. This taxonomy has been derived by analogy from the censor's classification used in evaluating censorship circumvention systems~\cite{Amir:SP:ParrotIsDead}. A \textit{local} warden controls at most a few network devices, is able to monitor only a small number of connections and its computational and/or storage resources are limited (it can be further divided into local host or local network warden). A \textit{state-level oblivious} warden has access to a vast number of devices spread geographically but it cannot keep network traces for a long time nor perform complex traffic analyses (due to limited computational and storage resources). It can perform, e.g.,\ deep-packet inspection but can inflict it on short observations of network traffic (single packets and not many connections). Finally, a \textit{state-level omniscient} warden controls geographically dispersed devices, can perform a variety of complex traffic analyses and store vast amounts of the intercepted traffic. When evaluating new information hiding techniques, this taxonomy allows to clearly indicate which type of adversary is considered in a threat model.

When it comes to the warden structure, it can monitor network traffic in a single (\textit{centralized} warden) or at multiple locations in the network at the same time and share gathered information among them (\textit{distributed} warden). In general, the localization and the number of locations in which the warden is able to inspect the traffic have a significant impact on the warden effectiveness.

\begin{figure}[!t]
\centering
\includegraphics[width=0.65\textwidth]{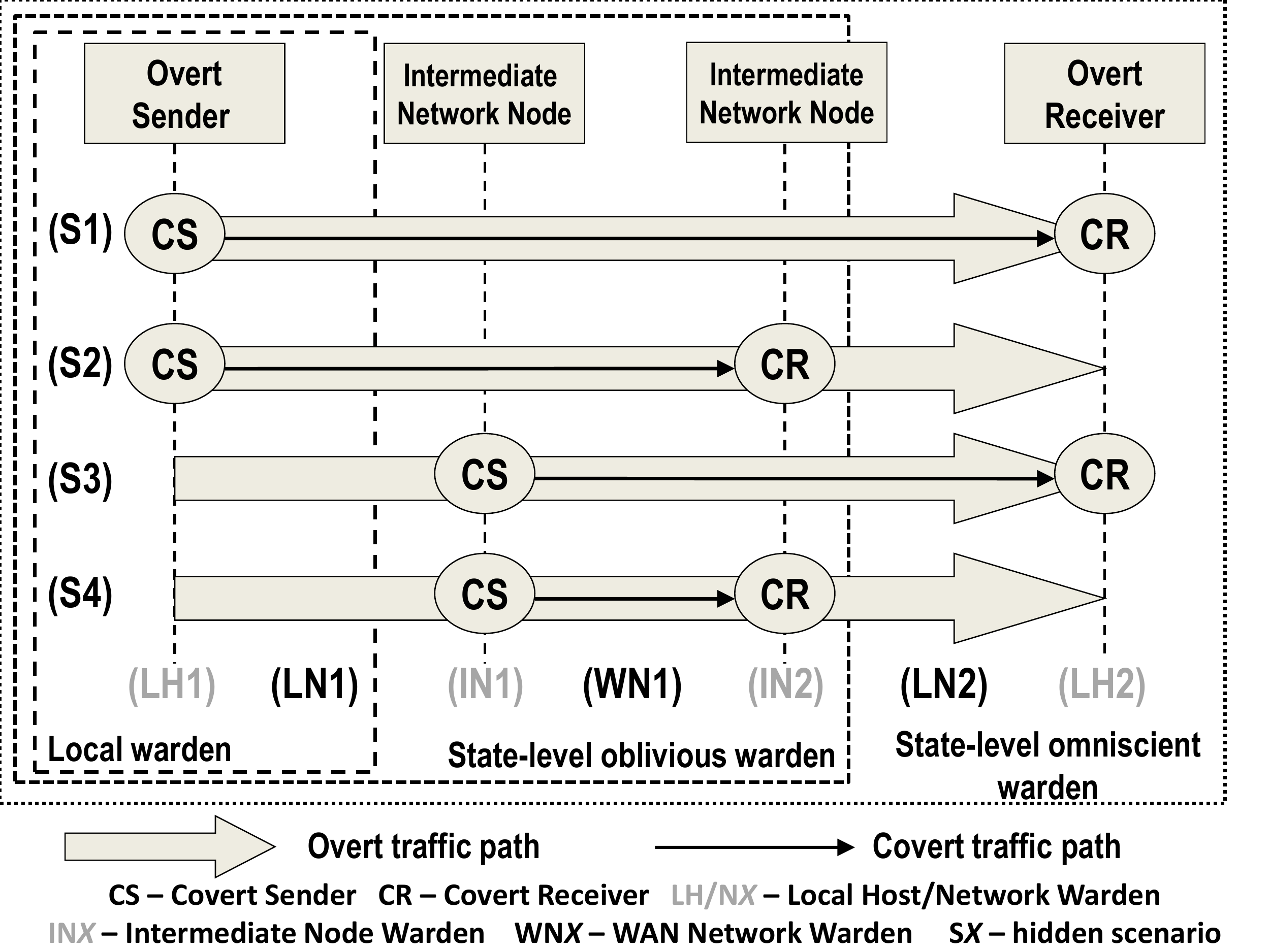}
\caption{Covert communication scenarios and potential warden structure and locations.}
\label{Fig:CovCommScen}
\end{figure}

The warden's localizations and structure are closely related also to the typical hidden communication scenarios (Fig.~\ref{Fig:CovCommScen}) as they are interdependent. Below we systematically reveal these relationships to better understand various types of wardens and how they can be realized to maximize the probability of successful network covert channel detection and prevention.

The possible hidden communication scenarios (S1 to S4 in Fig.~\ref{Fig:CovCommScen}) influence the passive and active strategies for the warden. Scenario S1 is the most common: the sender and the receiver establish a network connection while simultaneously exchanging secret data. The overt data communication path is the same as the hidden path. For the next three scenarios (S2 to S4 in Fig.~\ref{Fig:CovCommScen}), only a part of the end-to-end path is used for a hidden communication, as a result of actions undertaken by intermediate nodes (IN). Note that, in principle, in scenario S4, the overt sender and receiver are unaware of the covert communication's existence.

Considering the presented scenarios, three main localizations for a network warden are possible (denoted in Fig.~\ref{Fig:CovCommScen} as LN1/2 and WN1) and four localizations for a local host warden (denoted in Fig.~\ref{Fig:CovCommScen} as LH1/2 and IN1/2). The main difference between local host and network warden is that the former is also able to analyze dependencies between processes on the machine and link them to the traffic that is generated. This is especially important if on a certain host a covert sender is active and uses a network covert channel technique to impact the traffic, e.g.,\ by modifying header fields or by modulating the packets' timing. In this case, such anomalous behavior can be easily spotted. In contrast, network wardens mainly focus on analyzing traffic at the network level 
as
performing attribution at this point is sometimes hard to achieve.

Below, we describe the centralized/distributed warden case for the local and state-level oblivious wardens. Note that in Fig.~\ref{Fig:CovCommScen} borders between local, state-level oblivious and omniscient wardens are only contractual.

For the centralized local warden (typically this is the most realistic assumption for  real-life networks) two potential locations are possible: \textit{i)} the traffic generation and processing activities are performed on a local host (LH1) or \textit{ii)} the traffic exchanged between internal and outside networks is analyzed by the network warden (LN1). In the case of a distributed warden, information about the traffic can be exchanged and correlated between LH1 and LN1 and then any potential anomaly can be reported (such realization is similar to an IDS with distributed network and host sensors). A local warden is able to potentially detect the existence or eliminate a covert channel only in scenarios S1 and S2 as in the remaining scenarios the hidden data path is located outside its jurisdiction. 

In case of a centralized state-level oblivious warden, the warden is able to inspect processes and the generated traffic on a device located in the LAN (LH1) or WAN (IN1 or IN2). Alternatively, a network warden can be installed within a local area network (LN1) or in a wide area network (WN1). In general, it must be noted that centralized monitoring of the traffic at a single location (within a network or at host/node) offers only limited capabilities to detect covert communication. However, in case of IN1, IN2, and WN1, disturbing covert channels in all scenarios is possible. Alternatively, when a distributed state-level oblivious warden is considered then information gathered by each warden can be shared between the local host/intermediary node wardens (LH1, IN1/2) and network wardens (LN1, WN1). In result, more opportunities for the detection and targeted elimination are possible (blind elimination can be potentially applied at each node). However, it must be reminded that a state-level oblivious warden possesses only limited computational and storage capabilities, and thus eventually some information hiding techniques can remain undiscovered.

Finally, a state-level omniscient warden has a wider view on the network infrastructure as a local host or intermediary node warden. 
Also,
a network warden can be (practically) installed anywhere in the network. The considerations regarding its centralized and distributed structure are the same as in the previous case. However, a state-level omniscient warden has capabilities to perform complex processes and traffic analyses and store vast amounts of traffic and thus potentially offers more powerful detection opportunities.

\subsection{Warden Knowledge}
As already mentioned, in the first concepts of a network warden \cite{Fisk:EliminStegInternet}, it was assumed that no knowledge about its surroundings or possible information hiding techniques is available. However, later solutions \cite{Lewandowski:NAAW} started to consider additional knowledge which a warden can utilize in order to be more effective. In general, such knowledge can typically pertain to the network topology or specifics about a covert communication.

If we consider a stateful warden (passive or active), then it can be further classified into network-aware or network-unaware. A \textit{network-aware} warden has knowledge about the topology of a network in which it operates and implements traffic inspection~\cite{Lewandowski:NAAW}. This gives the warden an advantage as more information can be deduced about a covert communication from the traffic context. Obviously, a \textit{network-unaware} warden lacks this knowledge.

The knowledge about a covert communication is directly related to potential countermeasure types that a warden can apply. Kaur et al.~\cite{Kaur:ARES:CCIntContrProto} proposed a knowledge classification for wardens in the context of covert channel-internal control protocols (so-called \textit{micro protocols}~\cite{Backs2012}). They differentiate between four cases in which the warden is either unaware of a covert communication, aware of the presence of a covert channel, aware of the channel's coding and syntax, or fully aware of all related details of the covert channel. 
We adopt this concept and distinguish the knowledge of wardens in a slightly different way. The warden's knowledge can comprise the following aspects:
\begin{itemize}
\item Embedding and extraction processes of the network steganography methods (applicable for a local host warden):
\textit{(i)} the embedding process and micro protocol operation at the covert sender-side; \textit{(ii)} the extraction process and micro protocol operation at the covert receiver side.
\item The transferred covert channel packets between the sender and the receiver, i.e.,\ the representation of the steganographic technique in the overt communication channel. Such representation takes the form of a so-called \textit{hiding pattern}~\cite{Wendzel:CSUR:Patterns} and it describes a steganographic technique on a generic level. Such hiding patterns can be observed at the local or state-level network wardens.
\end{itemize}

Considering the above, based on the knowledge about the network covert channel that could have been potentially utilized for the hidden communication, warden attacks can be classified into four main types (by an analogy to \cite{Kaur:ARES:CCIntContrProto}). Obviously, attacks on covert channels can be performed only by active wardens.

Firstly, the warden can be unaware of the presence of the information hiding technique but suspects that some covert communication could be taking place. In such a case, a warden has no knowledge of the specifics of the information hiding technique. In result, only normalization rules based on the previous experiences can be applied in the hope that this will be sufficient to interfere or to destroy the covert channel. This type of attack is a form of a blind attack.

Secondly, the warden can be only aware of the presence of the steganographic method but it possesses no knowledge about the specifics of the technique used to create a covert channel. In this case, it tries to influence its operation and semantics randomly, e.g.,\ by injecting false data or erasing some packets, sending falsified commands to entities potentially involved in the covert communication (e.g.\ fake commands to terminate the covert communication in the hope to match the channel's syntax). Then it observes and reacts to potential anomalies in the responses. This, too, is a form of a blind attack.

Thirdly, the warden can be partially aware of the presence and the specifics of the information hiding technique used to perform covert communication. The knowledge that the warden has can be related to the syntax or semantics of the method and it can be efficiently utilized to inject targeted, legitimately appearing but falsified data to interfere with or to eliminate the covert channel. This type of attack is a form of a targeted attack.

Finally, the warden can be fully aware of the presence and specifics of the technique used, and even the hidden data scenario in which covert communication takes place. In this case, the warden can try targeted attacks to exploit weaknesses in the network covert channel design. If such weaknesses are identified, the warden has the largest possible control, i.e.,\ it can not only destroy the covert channel efficiently but it can also eavesdrop or influence it, e.g.,\ by impersonating one of the communication sides. This, too, is a form of a targeted attack.

\subsection{Warden Behavior}
Actions that the warden can perform in order to counter network covert channels include \cite{NIHBook}: \textit{i)} becoming aware of the channel's existence and documenting it; \textit{ii)} eliminating or limiting the covert channel's use; or \textit{iii)} detecting and auditing covert channels. Before any actions can be taken against an information hiding technique, it first needs to be identified, i.e.,\ the warden needs to become aware of the (potential) covert channel presence. Afterwards, the warden can eliminate the network covert channel or reduce its capacity below an acceptable limit (depending on the particular security policy). This is, as already mentioned, typically achieved by means of traffic normalization. Covert channels that cannot be eliminated or limited should be audited so that reliable detection methods are required for this purpose. For network covert channel detection, statistical approaches (e.g.~\cite{cabuk2004ip}) or machine learning techniques are often utilized. However, it must be noted that currently there is no one-size-fits-all solution available that will enable detection of all information hiding techniques. Typically, a specific detection method is crafted for each type of the covert channel.

Warden functionality concerns the inner mechanisms utilized which affects its efficiency and operations. As mentioned in Section~\ref{Sect:Introduction}, a warden can behave \textit{passively} or \textit{actively}. Both warden types can be also stateless or stateful. A \textit{stateless} warden does not take previous packets into account whereas a \textit{stateful} warden keeps historical information of previously analyzed packets and uses it to evaluate currently inspected traffic. Obviously, the amount of the historical data to be stored and processed is an important parameter as it impacts overall warden performance.

Considering the above, it is clearly visible that an active warden offers more opportunities from the countermeasures perspective. In more details, an active warden can also behave proactively or reactively. When it is \textit{proactive} it can intentionally issue `probes' (e.g.\ crafted packets) for known (or unknown) covert channels to provoke responses that will betray that the hidden communication is taking place. It can also assume in advance that its main aim is to interfere with any covert communication (and not to detect it) and induce certain rules (typically by means of normalization) that will affect all passing traffic whether it contains secret data or not (blind behavior). Alternatively, a warden can behave \textit{reactively} and trigger an action only when certain events or patterns indicate that the covert data exchange has been observed (targeted behavior). Obviously, a hybrid solution that combines both proactive and reactive behaviors is also possible.

Finally, what is most related to the content presented in the following sections of this paper, the warden can behave in a \textit{static} or in a \textit{dynamic} manner. Up till now practically all wardens functioning was static, i.e., their active normalization rules  remained constant in time or changed only slightly and rarely. This, as already mentioned, has certain disadvantages as the adaptive covert communication parties can infer the warden's normalization capabilities and adapt to them. Alternatively, as proposed in this paper, the warden can apply a dynamic operation strategy in which it performs periodic, randomized shuffling of the active subset of normalization rules. In result, the warden is able to evolve its normalization behavior over time which disturbs the functioning of the adaptive covert communication parties. The details on the dynamic warden behavior strategy are described in the next section.

\section{Dynamic Active Wardens}\label{Sect:DynWarden}

In this section, we introduce our concept of a dynamic warden. To this aim we first describe the underlying threat model, followed by a detailed description of the warden's inner workings.

It must be noted that if we take into account the warden taxonomy introduced in Fig.~\ref{Fig:Taxonomy}, the concept can be described as follows: the dynamic warden is an active, stateless warden with a proactive behavior. Its main aim is to eliminate or limit potential covert channels. It can be deployed as a centralized network warden that can be part of a local, state-level oblivious or omniscient defense. Specifically, it can be also unaware of the network topology and of the presence of a particular hidden data exchange.

\subsection{Threat Model}\label{Sect:NELPhase}
Yarochkin et al.\ first proposed an adaptive covert communication scenario~\cite{Yarochkin}. It consists of two parts: the \textit{network environment learning} (NEL) phase and \textit{communication} (COM) phase. During the NEL phase, the covert sender and the covert receiver are trying to identify the most suitable covert channel for performing their hidden communication in the presence of an active warden. To achieve this, the covert communication parties first passively monitor network traffic to infer information on the utilized network protocols. Based on this knowledge, they activate a handshake process, select suitable network protocols and send packet probes to each other in order to verify whether a chosen protocol will be blocked by the warden. This process continues unless at least one suitable protocol for a covert communication is determined. After the NEL phase is initially completed, the actual secret data exchange starts (the COM phase). However, the NEL process remains active to determine other suitable protocols and to discover potential changes in the administrative setups such as newly introduced warden rules that will block previously usable protocols.

In this paper, we assume that the number of available network covert channels which can be used by the adaptive covert communication parties is always (at least slightly) larger than the set of normalization rules available to the warden. This is a realistic assumption as it is commonly known that, considering the number of potential data hiding techniques and available network protocols, it is practically impossible to completely eliminate opportunities for covert communication~\cite{NIHBook}, \cite{Moskowitz1994}. In the case of a regular (static) warden, the adaptive covert sender and the covert receiver will eventually always find a way to exchange data in a hidden manner.

\begin{figure}[!t]
\centering
\includegraphics[width=0.65\textwidth]{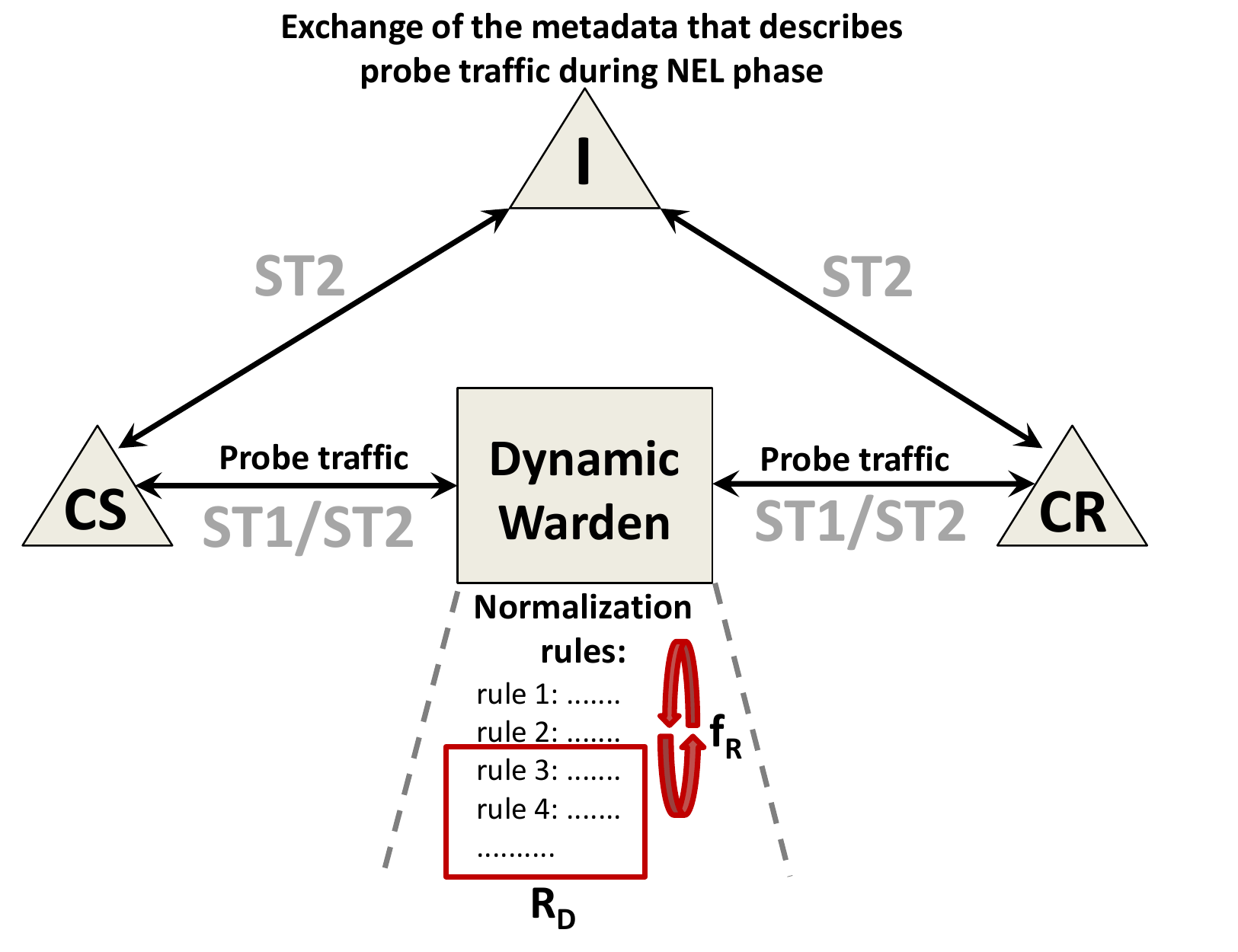}
\caption{Dynamic warden concept to defeat strategies performing a NEL phase.}
\label{Fig:DynWarden}
\end{figure}

Considering the above, the covert communication parties can use two strategies as introduced by Wendzel \cite{Wendzel:Sich12} to determine the active warden's ruleset (Fig.~\ref{Fig:DynWarden}): 
\begin{enumerate}
\item \textit{Direct Communication Strategy (ST1):} the covert sender (CS) and the covert receiver (CR) use a pre-defined packet sequence with different network protocols and types of bit manipulations in their protocol headers to infer if the active warden is present and which normalizing strategy it is using (ST1 in Fig.~\ref{Fig:DynWarden}). Such an approach improves over Yarochkin et al.~\cite{Yarochkin} as it utilizes an active behavior instead of only passively monitoring network traffic. In addition, this approach does not only analyze whether some network protocol is normalized by the warden or not, but also which bits of a protocol header undergo normalization~\cite{Wendzel:Sich12}.

\item \textit{Indirect Communication Strategy (ST2):} CS and CR use a third (usually temporary) participant IN (intermediate node) that is reachable by both CS and CR (ST2 in Fig.~\ref{Fig:DynWarden})\footnote{Please note that this scenario is also applied to construct covert channel overlay networks ~\cite{Szczypiorski2008}, \cite{Backs2012} where new links must be established between communicating covert nodes. The utilization of a (slow) temporary IN is a supporting factor in such scenarios~\cite{Wendzel:Sich12}.}. Using node IN, CS and CR are not able to perform covert communication to the full extent, e.g.,\ this covert channel has a very low capacity or can only be used temporarily. However, they are able to exchange via IN some limited meta-data regarding the probe traffic they try to send through the communication path ST1~\cite{Wendzel:Sich12}. In other words, they can announce probe traffic to each other and provide feedback about whether the probe packets were received successfully. This way, they are able to successfully determine which types of protocols can be utilized for the covert communication. It is worth noting that IN is not necessarily required to be aware of the covert communication between itself and CS or CR~\cite{Backs2012}. 
\end{enumerate}

Both mentioned approaches have their limitations, i.e.,\ the first strategy (sending a pre-defined sequence of probe packets) is error-prone and the second strategy depends on a third party IN connected to CS and CR. However, experiments that have been performed using popular normalizers~\cite{Wendzel:Sich12} have shown that normalizers can still be beaten especially by the second strategy. We conclude that both strategies are efficient even if they face a combination of all popular normalizers. It turned out that the first strategy can only be affected if a large part of the sequence sent is normalized and then the solution is to choose protocols not supported by the normalizers. Since a normalizer cannot affect an already working covert channel via a third party IN, the second strategy can be applied in any case.

Taking the threat of an adaptive covert communication scenario 
(continuous NEL phase with strategy ST2 in Fig.~\ref{Fig:DynWarden})
as a starting point, we want to reinforce the warden to be able to counter both offensive scenarios aimed at inferring warden capabilities. Therefore, we propose a new \textit{dynamic} strategy for an active warden. 
The proposed approach can be treated as a type of \textit{moving target defense} \cite{FinkEtAl:AntBased} and aims to increase the uncertainty of CS and CR about the normalizing strategy employed by the warden. This will be achieved by systematically and periodically modifying the warden's normalizing behavior over time. In other words, during the NEL phase or at any other time instant, inferring which protocols can be efficiently used for covert communication purposes will be an increased challenge for the covert communicating parties.

\subsection{Dynamic Warden Concept}

The dynamic strategy for an active warden is characterized as follows. We assume the presence of CS and CR that know that a normalizing active warden exists on their communication path and they try to devise a way to bypass it. Therefore, CS and CR utilize an adaptive covert communication system with the NEL and the COM phases as described in Section~\ref{Sect:NELPhase}. We further assume that the NEL phase can be performed in parallel to the covert data exchange. CS and CR try to infer the warden's active rules, i.e.,\ the way it works, and if they can correctly transfer covert data through the warden during the NEL phase, they will use these successful (i.e.\ non-blocked) covert channel techniques to exchange secret data during their COM phase.
If a typical \textit{regular} warden, i.e.,\ the one with a static set of applied normalization rules, is used, CS and CR will sooner or later establish a suitable covert channel to exchange secrets (by continuously switching between various potential network covert channels until at least a single non-blocked covert channel is found).
If we let $R_T$ (T for total) denote the 
set
of all possible rules that in an ideal case are able to counter all possible network covert channels, then as previously mentioned, we assume that a regular warden can have a set $R_R$ of rules which are a subset of $R_T$ ($|R_R| < |R_T|$).

In contrast, a \textit{dynamic} warden can randomly choose $R_D$ rules from $R_R$ ($|R_D| \ll |R_R|$) and apply these rules for $f_R$ seconds (reload interval). To counter the adaptive covert communication parties scenario, the dynamic warden applies $R_D$ normalization rules independently of the actual network traffic or any other outside information but in a constantly varying manner as after $f_R$ seconds another $R_D$ randomly chosen rules will be selected from $R_R$ and applied for the same period of time. The newly chosen rules are independent from the previously chosen rules. 
%
Thus, it will be difficult for the CS/CR to influence the warden's behavior in any way and this renders inferring suitable network covert channels between CS and CR a hard and time-consuming task as the subset of rules each time is randomized.

In result of the dynamic warden's actions, there is a situation as follows: CS and CR know that the warden exists and they launch the NEL phase, e.g.,\ through the node IN. After some time, they find a suitable protocol for their covert communication. Then, they initiate a hidden communication (COM phase). However, after a short time, the warden's rules are changed, making it harder (or, optimally, impossible) for the CS and CR to communicate any further. Finally, they have to select another suitable overt protocol for hidden communication purposes or reinitiate the NEL phase. This sequence of events, i.e.,\ finding a suitable covert channel, re-starting the covert communication, and ``loosing'' the previously usable network covert channel, continues in a loop. This way, CS and CR may be ``paralyzed'' or at least their communication is prolonged so much that in result the secret communication becomes more visible and less efficient.

\section{Experimental Evaluation}\label{Sect:Eval}
In this section we first describe the experimental test-bed and the proof-of-concept implementation of the dynamic warden in subsection \ref{testbed}, followed by the outline of the utilized experimental methodology in subsection \ref{meth} and presentation of the obtained results in subsection \ref{exps}. 

\subsection{Experimental Setup and Proof-of-Concept Implementation}
\label{testbed}
\begin{figure}[!t]
\centering
\includegraphics[width=0.65\textwidth]{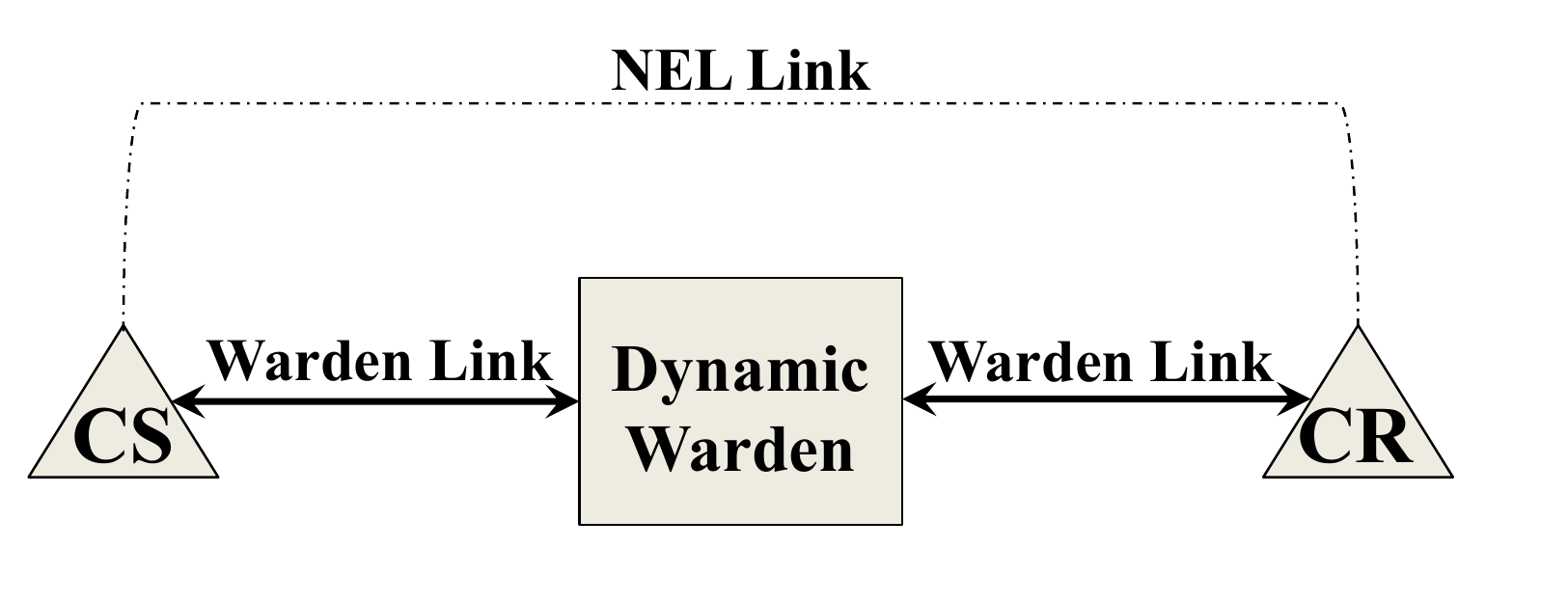}
\caption{Experimental setup for comparing the performance of regular and dynamic wardens.}
\label{Fig:ExpSetup}
\end{figure}

To experimentally evaluate our proof-of-concept implementation of the dynamic warden and to compare this concept with the regular warden, we created an experimental test-bed (Fig. \ref{Fig:ExpSetup}) consisting of three hosts (CS, CR and a warden). Each host has two network interfaces. As presented in Fig.~\ref{Fig:ExpSetup}, there are two distinct network communication links: the `Warden Link' between CS and CR and the `NEL Link' (which represents the connection over a temporary intermediate node IN). The `Warden Link' is used between CS and CR to perform the covert communication (COM phase) and to realize part of the NEL phase (send probe packets) through the warden. Each probe packet is sent by CS using the \textit{Scapy} tool and the packet received by CR is captured by \textit{libpcap} filters. Both adaptive communication sides can utilize one of 50 implemented network covert channels for every new packet they exchange. The `NEL Link' can be used between CS and CR only to exchange meta-data related to the sent and received probe packets. Please note that no intermediate node is required in this setup as the role of the intermediate node is only to forward packets, i.e.,\ we assume a ``perfect'' intermediate node by creating a direct NEL Link. 

As mentioned, our tool can exchange data over 50 different covert channels. These covert channels represent covert \emph{storage} channels that embed secret data into various areas of network protocol headers, e.g.,\ one covert channel encodes a secret information by setting the `reserved' flag of the IPv4 header, another covert channel sets the IPv4 `identifier' field to a pre-defined value (its presence indicates the secret bit), etc.

As a warden, we used a Python-based implementation that allows to influence network packets passing through the Linux netfilter firewall. For each covert channel technique there is one matching rule integrated into the warden to filter the particular covert channel. For instance, a rule to catch the above mentioned covert channel that embeds data in the `reserved' field of IPv4 is realized by checking for the presence of the set reserved bit. There is a 1-on-1 matching between covert channels and the set $R_T$ of filter rules, i.e.,\ each covert channel can be detected by exactly one rule of the warden.

In case of the \textit{regular} warden, we statically load our warden with a predefined subset $R_R$ of normalization rules. For the \textit{dynamic} warden, we randomly activate a selected subset $R_D$ of the filter rules for the interval $f_R$. After $f_R$ seconds the subset is shuffled and the next $|R_D|$ randomly chosen rules are selected. Both types of wardens have access to exactly the same set $R_T$ of normalization rules. For the dynamic warden the number of active rules and their activation time are configurable.

\noindent \textit{NEL phase details:} 
In our experimental setup, we let CS and CR perform a NEL phase as follows. 

First, CS announces probe packets (via the `NEL link') that will be sent to the CR. A single covert channel technique for this purpose is randomly selected. After announcement, CS transfers the particular probe packet through the 'Warden Link' while CR waits for the packet to arrive on its warden-connected interface. Each probe packet is sent five times in a row in order to alleviate the problem of packet loss occurrence. If one 
of the NEL phase's announced probe packets is successfully received by the CR (within a waiting time of $5\,s$), the particular covert channel technique is considered as non-blocked. 
The CR sends this information back to CS via the `NEL link'.
As soon as the CS and the CR establish at least one utilizable covert channel technique, they begin sending covert traffic, i.e.,\ they enter the COM phase.

However, if the CR does not receive the announced packet within a certain time period it will inform the CS (via the `NEL link') that the packet was filtered out, i.e.,\ the particular covert channel technique cannot be used to exchange data in a hidden manner.
Independently of the CR report, CS waits $1\,s$ and then starts to test another randomly selected covert channel type within the NEL phase. The set of non-blocked protocols is constantly updated by the NEL phase that is running in parallel to the COM phase. 

\noindent \textit{COM phase details:} 
As mentioned, when at least one non-blocked covert channel is determined by the CS and CR then the COM phase begins. In this case CS sends five packets with secret data using this particular covert channel type over the 'Warden Link'. Next, CS randomly selects another non-blocked data hiding technique that is available on the list of the non-blocked covert channels and 
sends
another five packets with secret data.
This process continues on and on. In case that there is only one non-blocked technique it will be continuously used. However, it must be noted that packets of the COM phase can be blocked by the dynamic warden whenever the ruleset is changed, i.e.,\ such packets may not reach the CR at all.

The CR uses a counter that is incremented whenever a COM phase packet is successfully received and we also count the number of normalized and forwarded (without modification) packets at the warden as well as we record the consumption of the CPU time and main memory consumption.
 
\subsection{Experimental Methodology}
\label{meth}
It must be noted that for the purpose of this paper we have decided to implement quite a sophisticated version of the NEL phase with a number of additional features of \cite{Wendzel:Sich12} when compared with original proposal from \cite{Yarochkin}. This includes, for example, an ability to switch between protocol field-specific (instead of only protocol-specific) covert channels or the ability to utilize the temporary intermediate node (IN) to circumvent the warden at least partially. Therefore such a realization of an adaptive network covert communication can be considered as a 
worst case scenario from the warden's perspective.

The experimental methodology for the dynamic and regular wardens is as follows. The measurement starts when the first protocol/covert channel announcement is received by the CR over the NEL link. Thus we measure the total time it takes to find the suitable covert channels and then to successfully transfer a certain number of packets with secret data (during the COM phase). It is worth noting that we deliberately do not count NEL phase packets as packets with covert information. Instead, we treat them as an ``overhead'' of the adaptive covert communication parties scenario. When CR obtains 
a pre-defined number of COM phase packets (e.g.\ 400), the measurement stops.
This rule is legitimate as CS and CR could use a fountain code such as LT code \cite{Luby2002} to encode their message: for a message comprising $p$ packets, CS can generate a potentially infinite stream of encoded packets, and CR can reconstruct (with overwhelming probability) the message from \emph{any} subset of encoded packets, where the size of the subset is only slightly larger than $p$ \cite{Luby2002}. Thus, CS that knows the message size, only needs to announce the number of required packets to CR via the NEL link.

\noindent\textit{Scenario Considerations:} During performed measurements we analyzed the following three scenarios:
\textit{(i)} the \textit{regular warden scenario} which features a static subset $R_R$ of rules that are active for the whole experiment; 
\textit{(ii)} the \textit{dynamic warden scenario} which periodically activates a smaller subset $R_D$ of randomly chosen rules (($|R_D| \ll |R_R|$)) for a short time period of length $f_R$;
\textit{(iii)} the \textit{no warden scenario} where the warden is inactive, i.e.,\ the warden is configured to just forward packets without any modifications.

For the regular warden, we experimented by activating a large subset of rules ($R_R$=95\% of all available rules\footnote{By $R_R$=95\%, we mean $|R_R|=0.95\cdot|R_T|$.}). The subset for the dynamic warden was significantly smaller: $R_D$ consists of 20\%, 30\%, or 40\% of all available rules. The collected results are averaged over the entire dataset of repeated trials to have a proper statistical relevance.

\subsection{Conducted Experiments}
\label{exps}
In this section we present results of the experimental evaluation of the dynamic warden concept. During our analyses we mainly focus on two aspects. First of all, we are interested to measure the dynamic warden \textit{effectiveness} in order to determine whether it is superior or inferior against the adaptive covert communication parties when compared with the regular warden. Secondly, we investigate the impact of the dynamic warden on the \textit{consumption of the resources} like CPU time and RAM. For the performed measurements we initially show results for the COM phase which consists of 400 packets with secret data. Later, we also analyze the impact of the length of data hiding exchange by comparing the scenario of 400 packets with 100 and 200 packets scenarios. Finally, we also investigate the completely random variants of the dynamic warden.

\subsubsection{Effectiveness Analysis}
\begin{figure}[!t]
\centering
\includegraphics[width=0.75\textwidth]{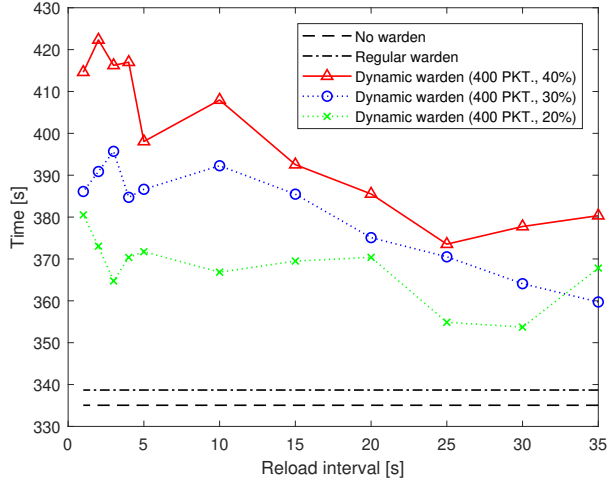}
\caption{Influence of the $f_R$ on the time needed to complete the transfer of 400 covert packets for different types of wardens.}
\label{Fig:DWresults1}
\end{figure}

In order to investigate the effectiveness of the dynamic warden, we first decided to determine the impact of the ruleset reload interval $f_R$ on the time needed by the adaptive covert communication parties to covertly transfer 400 packets with secret data. Obviously, the longer the time needed to complete such a hidden data transfer the better the warden performed (so the higher its effectiveness) as in result it is harder for the covert communication parties to establish a non-blocked covert channel and transfer secret messages. To determine the nature of this effect, we experimented with the different dynamic ruleset sizes ($R_D=$ 20\%, 30\%, and 40\%) for the dynamic warden and various reload intervals (from $1$ to $35\,s$). The obtained results are presented in Fig.~\ref{Fig:DWresults1}. 

First it must be observed that the regular warden with 95\% of continuously active rules is very ineffective as it makes the covert connection longer only by 1\% when compared with the `no warden' scenario. This shows that indeed in order to deal with the adaptive covert communication parties' threat other more effective solutions are desired. 

For the dynamic warden, we observe that the best results were yielded for the shorter reload intervals, i.e.,\ $2\,s$ and $R_D=$40\%. This means that the results were ca.\ 25\% better (with standard deviation of 19.05 s) when compared with the time needed to complete the covert data transfer for the regular warden scenario. Moreover, for all investigated reload intervals and all considered active ruleset sizes the dynamic warden performed better than the regular one. It must be also observed that the longer the reload interval is, the shorter is the time needed to complete the transfer of 400 packets with hidden data and the best results are achieved for the time interval in the range 1 to 5$\,s$. This follows intuition as the more often the ruleset changes the greater the chance that the list of non-blocked covert channels maintained by the adaptive covert communication parties is (at least partially) outdated.

\begin{figure}[!t]
\centering
\includegraphics[width=0.75\textwidth]{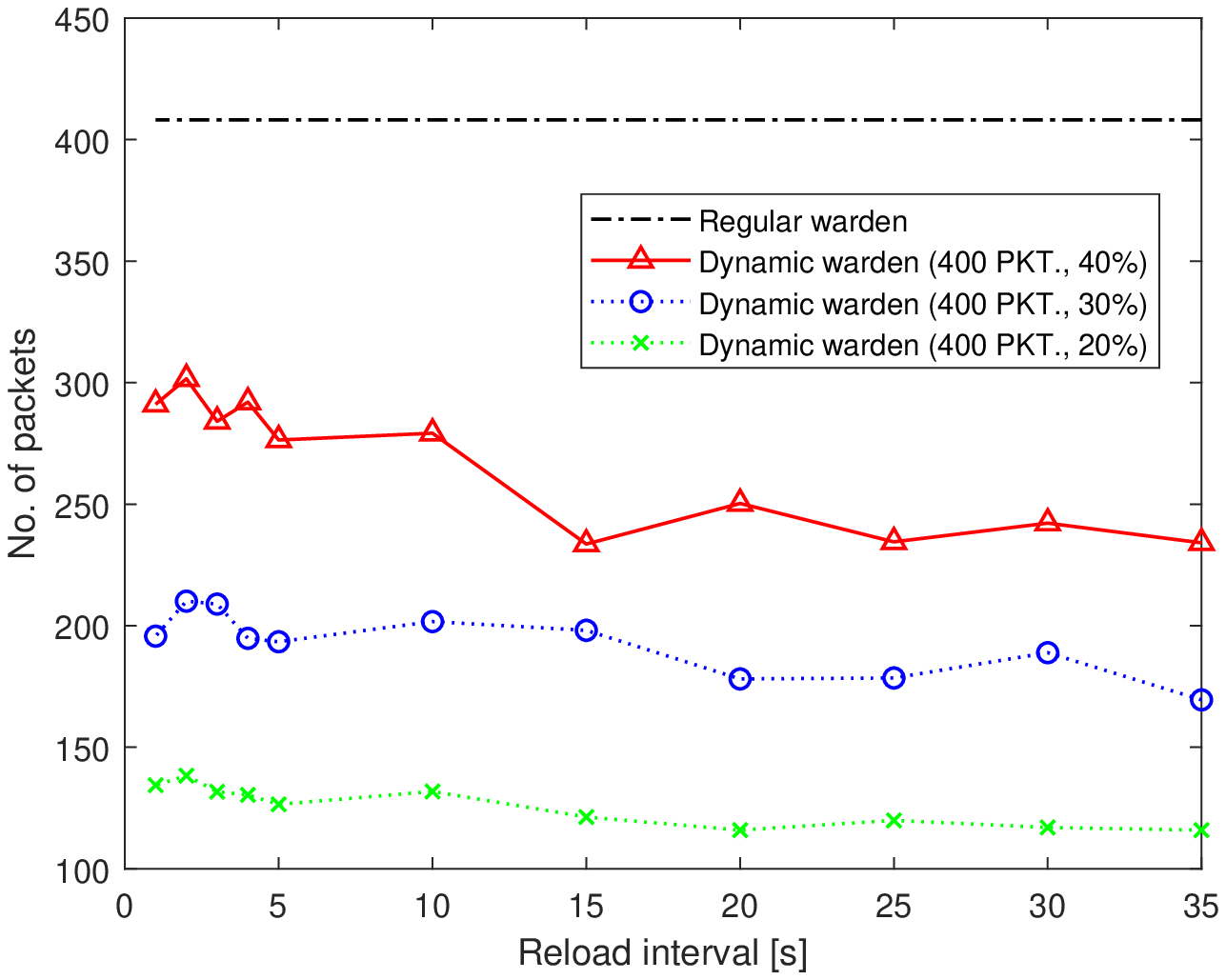}
\caption{Influence of the $f_R$ on the number of \textit{normalized} packets from CS to CR for different types of wardens.}
\label{Fig:DWresults4}
\end{figure}

\begin{figure}[!t]
\centering
\includegraphics[width=0.75\textwidth]{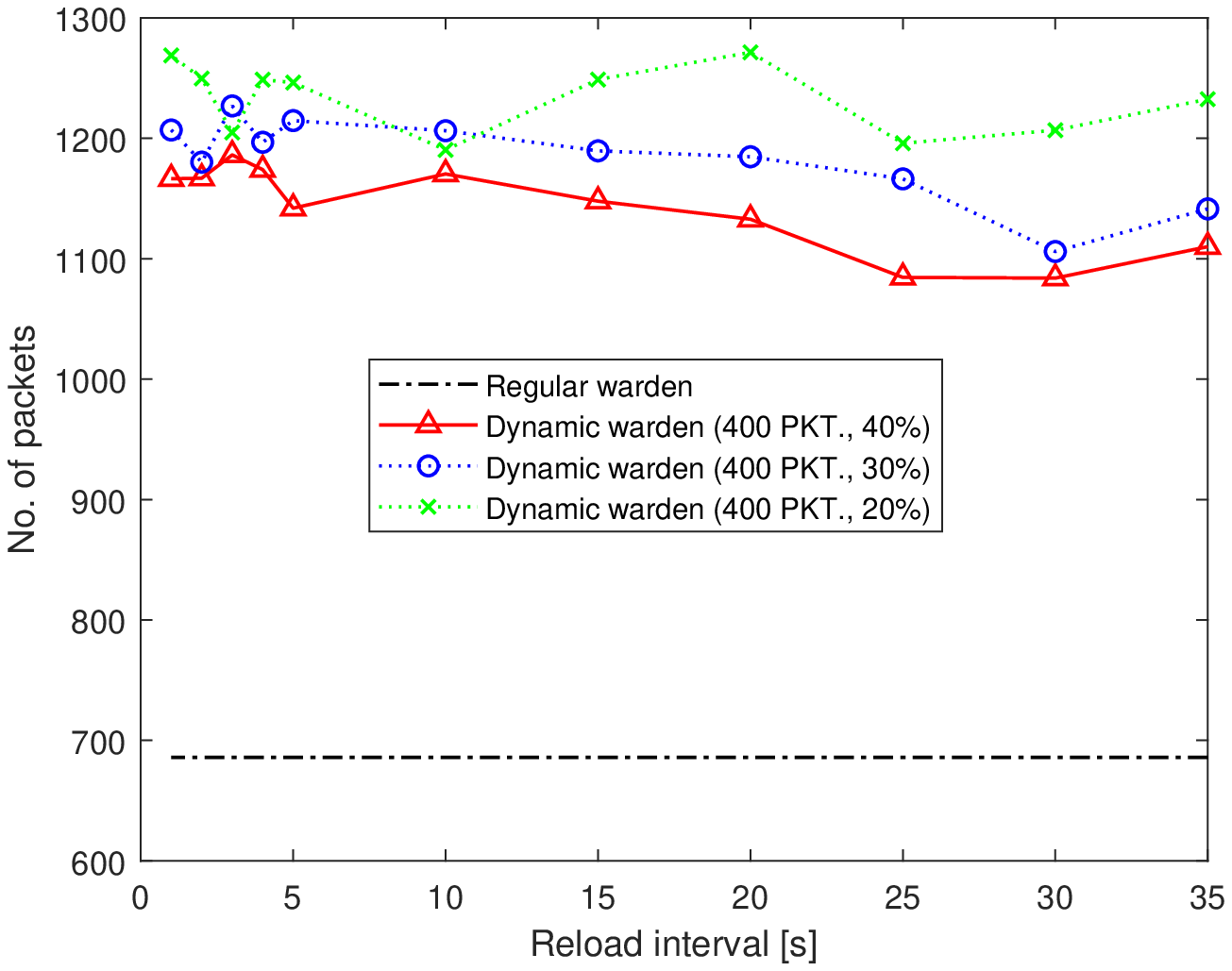}
\caption{Influence of the $f_R$ on the number of \textit{forwarded} packets from CS to CR for different types of wardens.}
\label{Fig:DWresults5}
\end{figure}

\begin{figure}[!t]
\centering
\includegraphics[width=0.75\textwidth]{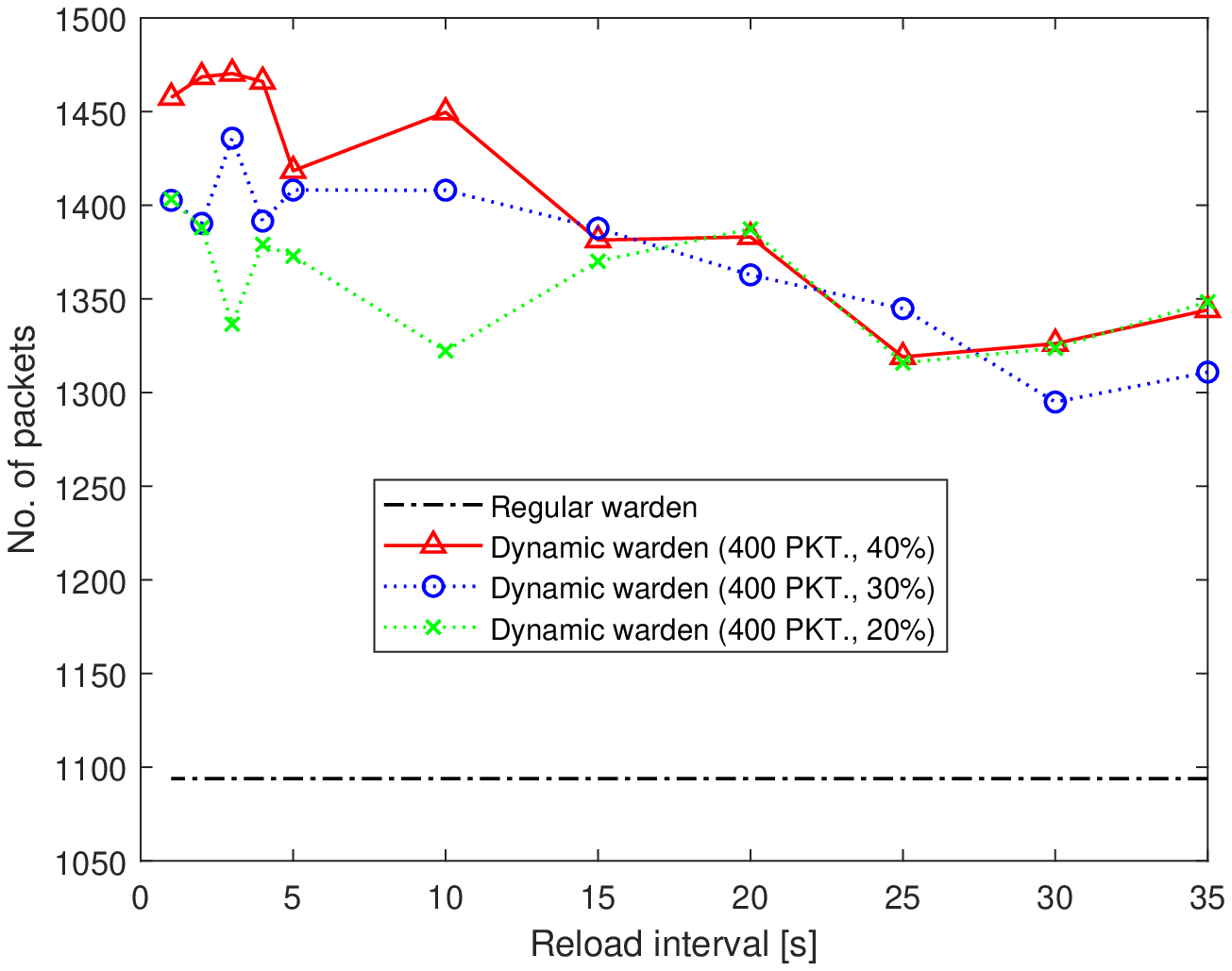}
\caption{Influence of the $f_R$ on the \textit{total number} of packets from CS to CR for different types of wardens.}
\label{Fig:DWresults6}
\end{figure}

Then, we explored the traffic statistics for both the dynamic and the regular wardens, i.e.,\ how many packets were normalized, forwarded without modification and exchanged in total. This allows to establish the cause why the dynamic warden outperforms the regular warden in the adaptive covert communication scenario.

The obtained results are presented in Figs.~\ref{Fig:DWresults4}, \ref{Fig:DWresults5}, and \ref{Fig:DWresults6}. Based on these figures it can be observed that although the number of normalized packets is lower for the dynamic warden by on average ca.\ 30\% for $R_D = 40\%$ (by ca.\ 50\% for $R_D = 30\%$ and by ca.\ 70\% for $R_D = 20\%$) the number of forwarded packets is much higher and in result the total number of exchanged packets is much higher for the dynamic warden than for the regular one (i.e.\ up to 35\% for $R_D = 40\%$, up to 30\% for $R_D = 30\%$ and up to 25\% for $R_D = 20\%$). 

This explains why the total time needed to complete the transfer of 400 covert packets under the dynamic warden is much longer than for the regular warden. Due to the periodical changes in the active ruleset the adaptive covert sender must continuously modify the covert channel used with increasing uncertainty.
Although a covert channel may have previously been identified as non-blocked, a moment later the active ruleset on the dynamic warden changed, resulting in a blocked covert channel and the need to transfer additional probe packets (NEL phase). It must be also emphasized that the prolonged covert transmission time together with the significantly higher number of packets that need to be generated and transmitted may allow additionally deployed detection measures to spot increased volume of traffic.

\subsubsection{Resource Consumption Analysis and Impact on Legitimate Users' Traffic}

In the next step, we investigated which impact each type of warden has on the memory and processor usage. 

\begin{figure}[!t]
\centering
\includegraphics[width=0.75\textwidth]{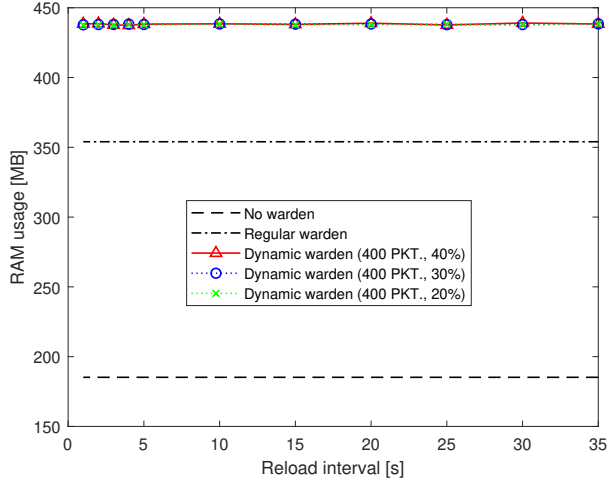}
\caption{Influence of the $f_R$ on the RAM usage for different types of wardens.}
\label{Fig:DWresults7}
\end{figure}

\begin{figure}[!t]
\centering
\includegraphics[width=0.75\textwidth]{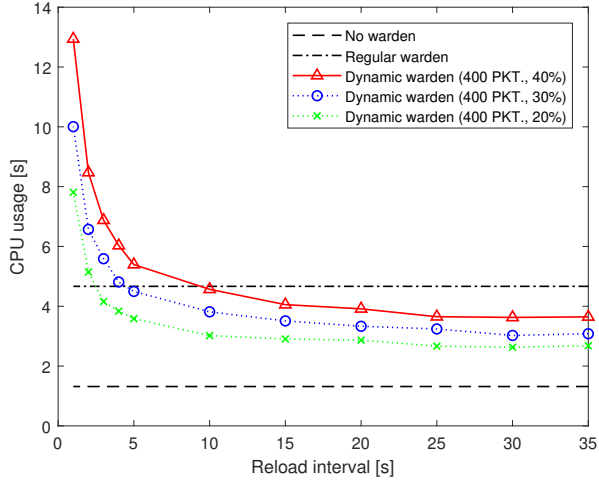}
\caption{Influence of the $f_R$ on the CPU usage for different types of wardens.}
\label{Fig:DWresults8}
\end{figure}

The obtained experimental results are presented in Figs.~\ref{Fig:DWresults7} and \ref{Fig:DWresults8}. The RAM usage seems to be not affected by the reload interval $f_R$ and it is about 25\% higher than in case of the regular warden (with a very low standard deviation of 0.25 MB).

As to the CPU usage, for the lowest $f_R$ values the load is the highest, e.g.,\ for $1\,s$ the load is ca.\ 280\% higher for $R_D = 40\%$ when compared to the regular warden (with a standard deviation of 0.6). This is understandable as shuffling the active ruleset for the dynamic warden so often must be reflected in the increased CPU consumption. However, it should be also emphasized that for slightly higher values of $f_R$, e.g.,\ $f_R=4 s$, the CPU consumption is the same for the dynamic warden with $R_D=$ 30\% and for the regular warden ($R_R=$ 95\%). This shows that it is still possible to find a suitable trade-off between the CPU consumption (which may be similar to those caused by the regular warden) and the effectiveness of the dynamic warden. For instance, for the dynamic warden with $f_R=10\,s$ and $R_D = 40\%$ the CPU consumption will be comparable while the time to complete the covert transmission will be increased by 20\%.

It must be also emphasized that with the link bandwidth growing beyond 1 Gb/s, and average TCP packet sizes of less than 1 kbit, a normalizer must handle more than 1 million packets per second.
Even if applying a rule needs only 100 instructions, a core running at 2~GHz and with an IPC\footnote{Instructions per cycle, even a superscalar core normally does not reach higher counts.} of 1.5 cannot apply more than 30 rules.\footnote{While the above figures are synthetic, they match projections of Handley et al.~\cite{Handley01} into current performance range.}
While the throughput can be increased by using multiple cores for different packets, latency still will grow beyond 1 microsecond for more than 30 rules, as fine-grained parallelism by using multiple cores for different rules on a single packet will not pay off because of parallelization overhead. Thus, negative side-effects for packets with real-time requirements, such as Voice-over-IP traffic, would be the consequence.
At the same time, the number of possible network covert channels still grows, which means that additional rules will be necessary if all covert channels are to be blocked.
Hence, an approach achieving this goal without using all rules all the time is necessary. 

It must be noted that results presented above are in line with the existing works discussing the security solutions' overhead in terms of resource consumption, e.g., \cite{Caviglione2012}, \cite{Merlo2015}.


Finally, we have also conducted experiments in order to determine the performance losses perceived by the legitimate users when the dynamic warden is deployed and compare the obtained results with the regular warden scenario. In order to evaluate this impact we conducted the following experiment. Due to its popularity we decided to utilize HTTP-based traffic, i.e., downloading of a 100MB file from an HTTP server to an HTTP client through the warden. As an HTTP server we used \textit{SimpleHTTPServer}\footnote{\url{https://docs.python.org/2/library/simplehttpserver.html\#module-SimpleHTTPServer}} version 0.6 and as HTTP client \textit{wget}\footnote{\url{https://www.gnu.org/software/wget/}} version 1.18. We measured the downloading time of the 100MB file first for the regular (static) warden (with 95\% active normalization rules) and then for the dynamic warden (with 40\% active normalization rules and 10 seconds reload interval -- we selected the configuration with the highest percentage of active rules, i.e., a worst case scenario). Each experiment has been repeated ten times to have a proper statistical relevance. From the obtained results we conclude that when the dynamic warden is deployed between the HTTP client and HTTP server the download time is ca.\ 7\% shorter than in case  of the regular warden scenario. This shows that the performance losses perceived by the legitimate users are lower for the dynamic warden setup.

\subsubsection{Impact of the Covert Transmission Length}

\begin{figure}[!t]
\centering
\includegraphics[width=0.75\textwidth]{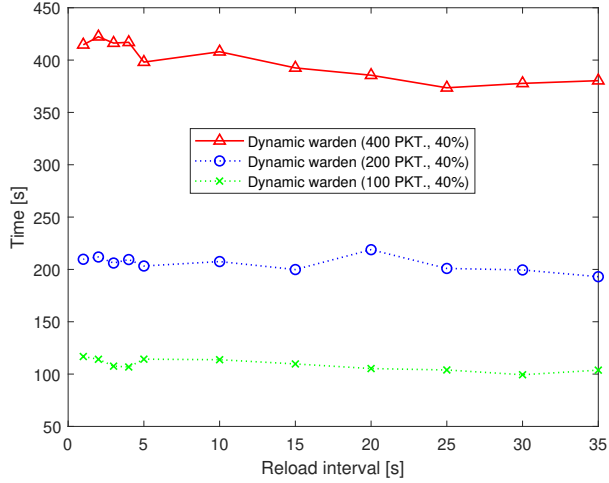}
\caption{Influence of the $f_R$ on the time needed to complete the transfer of covert packets for different lengths of the covert transmissions ($R_D=$ 40\%).}
\label{Fig:DWresults9}
\end{figure}

\begin{figure}[!t]
\centering
\includegraphics[width=0.75\textwidth]{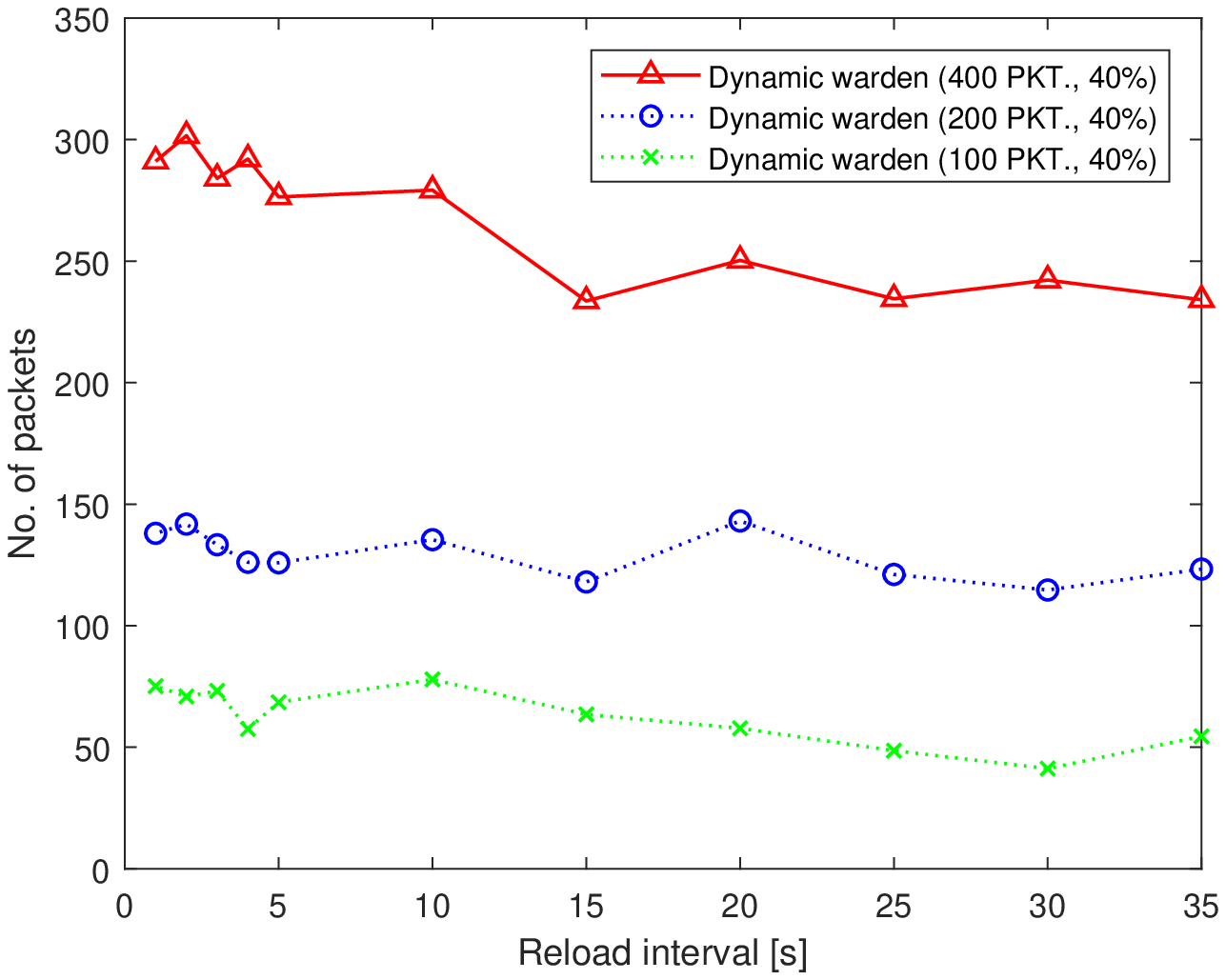}
\caption{Influence of the $f_R$ on the number of normalized packets for different lengths of the covert transmissions ($R_D=$ 40\%).}
\label{Fig:DWresults10}
\end{figure}

\begin{figure}[!t]
\centering
\includegraphics[width=0.75\textwidth]{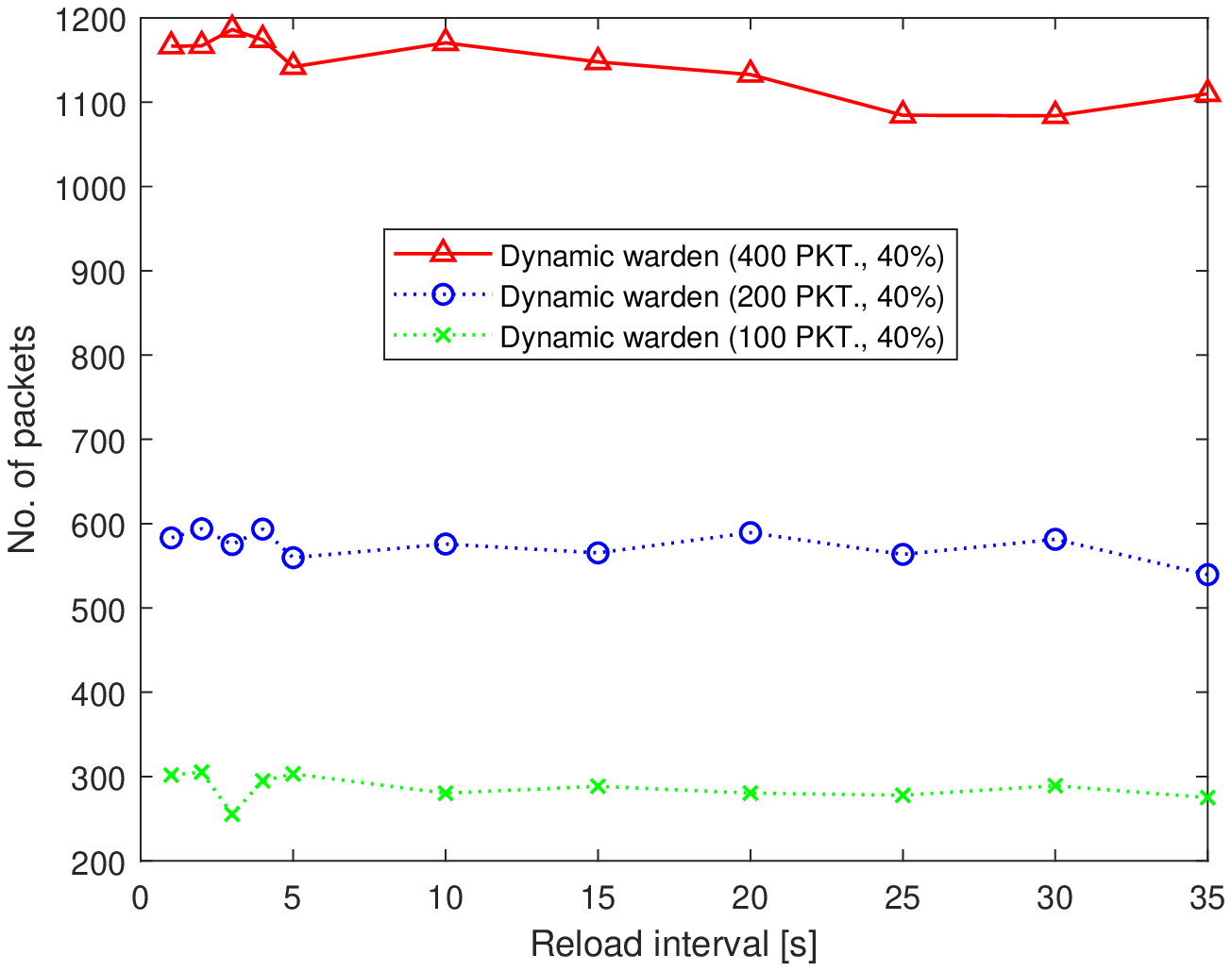}
\caption{Influence of the $f_R$ on the number of forwarded packets for different lengths of the covert transmissions ($R_D=$ 40\%).}
\label{Fig:DWresults11}
\end{figure}

\begin{figure}[!t]
\centering
\includegraphics[width=0.75\textwidth]{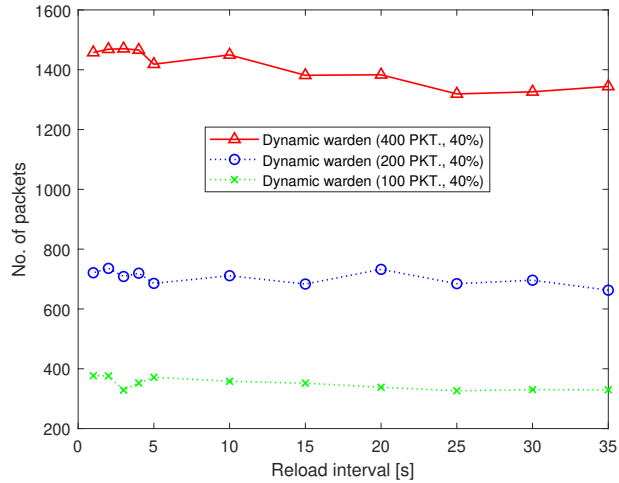}
\caption{Influence of the $f_R$ on the total number of packets for different lengths of the covert transmissions ($R_D=$ 40\%).}
\label{Fig:DWresults12}
\end{figure}

\begin{figure}[!t]%
    \centering
    \subfloat[]{{\includegraphics[width=0.50\textwidth]{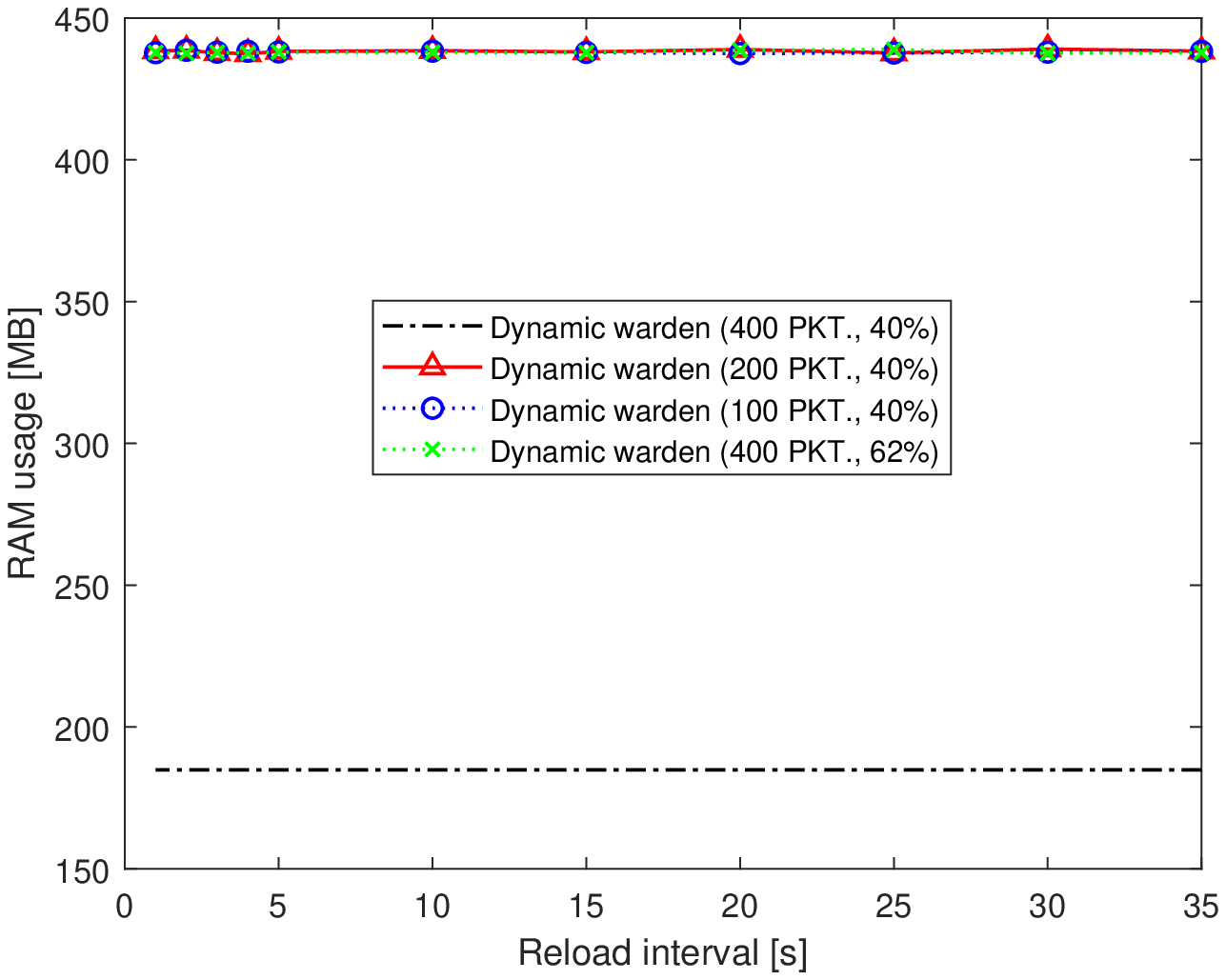} }}%
    \hspace{-0.5cm}
    \subfloat[]{{\includegraphics[width=0.50\textwidth]{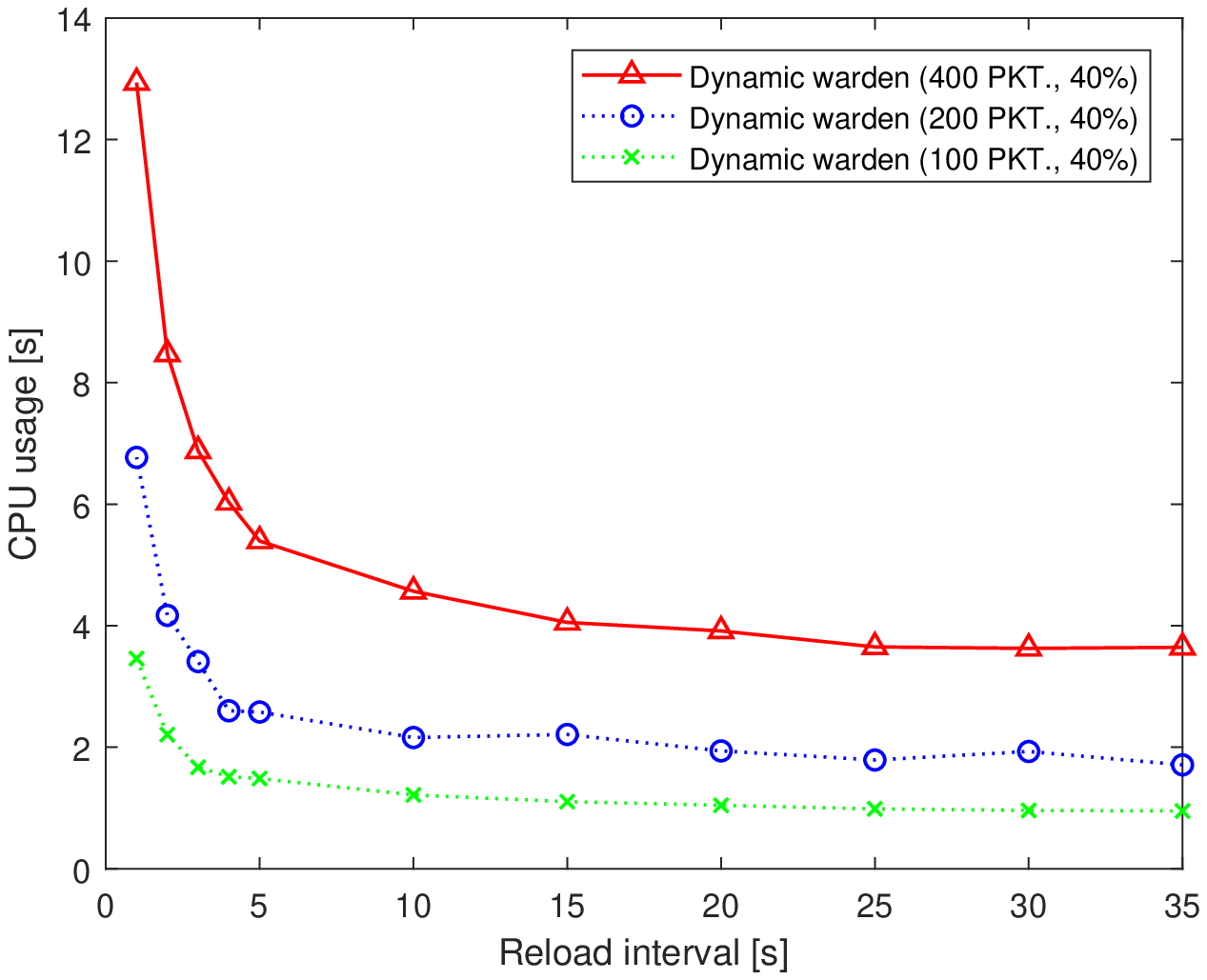} }}%
    \caption{(a) Influence of the $f_R$ on the RAM usage for different lengths of the covert transmissions ($R_D=$ 40\%). b) Influence of the $f_R$ on the CPU usage for different lengths of the covert transmissions ($R_D=$ 40\%).}%
    \label{Fig:13_and_14}%
\end{figure}

Finally, we also want to investigate how the length of the covert communication impacts the effectiveness and performance of the dynamic warden. For this purpose we varied the number of packets that need to be exchanged during data hiding transmission in the range 100 to 400. The obtained results are presented in Figs.~\ref{Fig:DWresults9} to \ref{Fig:13_and_14}. We restricted performed experiments to a dynamic warden with $R_D=40\%$ as it is the strongest among the three variants investigated. The results show that apart from the RAM consumption which is stable regardless of the length of the covert transmission the rest of the parameters change linearly. For example, the average time needed to covertly transfer 100 packets is ca.\ 100$\,s$, for 200 packets it is around 200$\,s$ while for 400 packets it is ca.\ 400$\,s$. Similar relationships are visible when taking into account the number of packets or CPU consumption.


%

\subsubsection{Random Dynamic Warden Variant}
Finally, we have also decided to investigate a \textit{random} variant of the dynamic warden. A random variant means that both $f_R$ and $R_D$ are selected randomly with equidistribution from certain ranges. We have considered the following four subvariants:
\begin{itemize}
\item V1: $f_R \in \langle 1s; 35s \rangle$ and $R_D \in \langle2\%; 100\%\rangle$ (i.e.\ between 1 and 50 rules). This means that the reload interval and the size of an active ruleset are selected randomly for the typical values investigated for the dynamic warden in the previous experiments.

\item V2: $f_R \in \langle 1s; 35s \rangle$ and $R_D \in \langle20\%; 40\%\rangle$ (i.e.\ the  size of the active ruleset is randomly selected between 10 and 20 rules). Such values were tested for the dynamic warden in the previous sections. 

\item V3: $f_R \in \langle 1s; 10s \rangle$ and $R_D \in \langle20\%; 100\%\rangle$ (i.e.\ between 10 and 50 rules). This means that the reload interval is selected from the values for which the best results have been achieved for the dynamic warden in the previous experiments. 

\item V4: $f_R \in \langle 1s; 10s \rangle$ and $R_D \in \langle20\%; 40\%\rangle$ (i.e.\ between 10 and 20 rules) -- both the reload interval and the size of the active ruleset are selected randomly in the ranges for which the best experimental results have been obtained for the dynamic warden investigated in the previous experiments.
\end{itemize}

The obtained results for all variants are presented in Table \ref{random}. The collected results are averaged over the entire dataset of 20 repeated trials to have a proper statistical relevance. 

From the obtained results it may be observed that the V3 variant offers the best results in terms of the time needed to complete the covert transfer and the volume of traffic generated by the adaptive covert channel parties which is comparable with the best results obtained in the previous subsections while offering lower CPU and RAM consumption. The second best variant is V1 with 3.6\% decrease in the time needed to complete data hiding exchange, while V2 and V4 results are similar and worse by 9\% when compared with V3.

Therefore the random variant must be considered also as a promising realization of the dynamic warden.

\begin{table}[]
\caption{Experimental results for various variants of the random dynamic warden.}
\tiny
\begin{tabular}{@{}cccccccccccc@{}}
\toprule
\begin{tabular}[c]{@{}c@{}}Var.\end{tabular} & \begin{tabular}[c]{@{}c@{}}\# of COM \\ phase pkts.\end{tabular} & \begin{tabular}[c]{@{}c@{}}Time of\\covert \\transfer {[}s{]}\end{tabular} & \begin{tabular}[c]{@{}c@{}}Std.\\ dev.\end{tabular} & \begin{tabular}[c]{@{}c@{}}CPU\\ usage \\ {[}s{]}\end{tabular} & \begin{tabular}[c]{@{}c@{}}Std.\\ dev.\end{tabular} & \begin{tabular}[c]{@{}c@{}}RAM\\ usage \\ {[}MB{]}\end{tabular} & \begin{tabular}[c]{@{}c@{}}Std.\\ dev.\end{tabular} & \begin{tabular}[c]{@{}c@{}}\# of \\ norm.\\ pkts.\end{tabular} & \begin{tabular}[c]{@{}c@{}}Std.\\ dev.\end{tabular} & \begin{tabular}[c]{@{}c@{}}\# of\\ forw.\\ pkts.\end{tabular} & \begin{tabular}[c]{@{}c@{}}Std.\\ dev.\end{tabular} \\ \midrule
\multirow{3}{*}{V1} & 400 &  400.2	& 21.71 & 4.17	& 0.4 &  354.09 & 0.85 &  319.25 &	43.8
 & 1070.5 & 92.8 \\
 & 200 &  207.16 & 21.05 &  2.16 & 0.3 & 354.02 & 0.92 & 159.95 & 27.43 & 535.4 & 62.32 \\
 & 100 &  106.88 &	16.12 & 1.11 & 0.33 & 353.57 & 0.49 & 77.5	& 28.28 & 252.35 & 34.14 \\
 \midrule
\multirow{3}{*}{V2} & 400 &  378.59 &	22.98  &  3.23 & 0.2 &  354.25 & 0.89 & 191.75	& 26.43 & 1177.8 &	83.88 \\
 & 200 &  192.35 & 14.57 & 1.62	& 0.11 & 354.01	& 0.75 & 96.55 & 11.41 & 578.1 & 42.02  \\
 & 100 &  103.32 & 11.79 & 0.87 & 0.1 & 354.1 & 1.1 & 43.35 & 11.72 & 297.9 & 38.06 \\
 \midrule
\multirow{3}{*}{V3} & 400 &  415.9	& 28.94  &  6.72 &	0.48 &  354.36 & 0.79 &  418.95 &	27.90 &  999.80 & 78.57 \\
 & 200 &  215.79 & 25.32 &  3.45 & 0.39 & 353.99 & 0.71 & 199.6	& 21 & 520.1 & 76.92 \\
 & 100 &  112.78 & 12.57 & 1.11 & 0.13 & 353.82 & 0.82 & 92.5 & 20.58 & 47.1 & 19.06 \\
 \midrule
\multirow{3}{*}{V4} & 400 & 379.29 & 18.22 & 4.15 & 0.19 &  354.07 & 0.69 &  199.55 & 23.77 &  1167.65 & 50.24 \\
 & 200 &  202.69 & 14.23 & 2.17	& 0.15 & 353.84 & 0.67 & 106.95 & 14.18 & 602.25& 35.81 \\
 & 100 &  107.31 & 13.79 & 1.11 & 0.13 & 353.79 & 0.61 & 44.05 & 13.7 & 305.45 & 39.37 \\ \cmidrule(l){1-12} 
\end{tabular}
\label{random}
\end{table}

\subsubsection{Summary}
\label{sum}
The dynamic warden turned out to be a more promising countermeasure against the adaptive covert communication parties than currently available regular (active) wardens as the time needed to complete the hidden data transfer for the adaptive scenario is significantly longer. Moreover, the dynamic warden ``forces'' the adaptive covert communication parties to exchange many additional packets in comparison to the regular warden. For example, in the case when 400 covert packets need to be transferred in the presence of the dynamic warden with $R_D=$ 40\%, ca.\ 1100 packets are needed in case of the regular warden and for the dynamic warden it is ca.\ 1450 packets (up to 35\% increase). In other words the dynamic warden ``buys'' more time for the defensive solutions to spot covert communication as well as it makes it more visible (due to the elevated number of packets needed to be exchanged by the adaptive covert communication parties). Thus, the dynamic warden slows down covert data exfiltration attacks. Especially, one solution is favorable when $R_D = 40\%$ and $f_R = 2\,s$. However, in this case the CPU usage is elevated which is a result of the frequent changes applied to the normalization table. If the high CPU consumption is an issue then the dynamic warden with $R_D = 40\%$ and $f_R = 10\,s$ should be selected as it has practically the same performance as a regular warden while at the same time prolonging the length of the covert transmission by 25\%.

Another alternative if the high CPU consumption is an issue is to realize the dynamic warden in a random variant where the CPU and RAM utilization are lower yet the other efficacy parameters are similar. The best variant of the random dynamic warden turned out to be the one where $f_R$ has quite a low value (i.e.\ randomly selected from the range $\langle 1s; 10s \rangle$) and $R_D$ is chosen from $\langle20\%; 100\%\rangle$ range. 
In this case the CPU consumption is lower by ca.\ 50\%, while the RAM usage decreases by ca.\ 25\% and is similar to the regular warden scenario.

As already mentioned our approach introduces random shuffling of normalization rules to counter the frequent change of the network covert channels used by the covert sender and the covert receiver. 
It must be noted that when CS/CR infer this behavior
they may further try to modify (simplify) their strategy, however, this will not be an advantageous option. Consider a simple scenario where CS/CR use a single covert channel (selected from the pool of available data hiding techniques) for the whole length of the covert transmission. In this case it does not matter which covert channel CS/CR will use as each of the data hiding techniques is only blocked part of the time, on average: $|R_D|$/$|R_T|$ (or 40\% of the time for the variant that turned out to be most effective for the dynamic warden).

Thus, the packets in which secret data is hidden will be successfully transferred in a fraction $1-|R_D|/|R_T|$ or 60\% of the time. This multiplies the time until a given number of covert channel-related packets is transmitted (compared to the no warden scenario) by $1/(1-|R_D|/|R_T|)$ or 1/60\%=1.66. 
This means that in this case the time needed to complete covert transmission would be 66\% longer (compared to 25\% increase in transmission time obtained in our experiments). This means that a dynamic warden cannot be easily overcome when the CS and CR are utilizing such a strategy (i.e. no change of used covert channel throughout the hidden data transmission).

\section{Conclusion and Future Work}\label{Sect:Concl}

Covert channels are an increasingly important factor within the cyber economy. Their presence influences the lifespan and quality of illegal and legal products, such as botnets, anti-malware software or data leakage (protection). In recent years, covert channels became more adaptive and sophisticated, resulting in a demand for improved countermeasures, i.e.,\ wardens.

In this paper, we first introduced a novel taxonomy that allows to comprehensively describe and understand wardens' features. Moreover, we identified weaknesses of existing regular (active but static) warden approaches, i.e.,\ that they are not able to efficiently interfere with an adaptive covert communication parties scenario.
In order to address this issue, we introduced a new type of warden, called a \textit{dynamic warden} which is able to effectively limit the capabilities of the adaptive covert communication parties and renders it difficult for them to infer the warden's normalization strategy. To prove that the proposed solution is effective, we
implemented a proof-of-concept and 
evaluated our dynamic warden concept by comparing it with a regular warden.
%
We demonstrated that a significantly smaller number of active normalization rules is needed to interfere with the adaptive covert communication parties scenario, i.e.,\ the data hiding transmission lasts up to 25\% longer and up to 35\% more packets must be exchanged. This provides defenders with an opportunity to detect and prevent further covert transfer as it is exposed for a longer time with a higher volume of data.

In currently ongoing work, we are going to implement a stateful version of the dynamic warden, which we call an \textit{adaptive} warden. The adaptive warden introduces normalization rules based on the previously seen traffic. The main goal of the adaptive warden is to require fewer rules than the dynamic warden for the same degree of covert channel limitation while providing a better performance (as fewer rules must be checked per packet).

Our future work will comprise a stochastic model of the dynamic (and adaptive) wardens with Markov chains to see how good the wardens exploit their possibilities, and what their limits are.

\section*{References}
\bibliography{bare_conf}

\end{document}